\documentclass[a4paper,11pt]{article}
\DeclareUnicodeCharacter{2212}{\ensuremath{-}}
\pdfoutput=1 
\usepackage{jcappub} 
\usepackage{aas_macros}
\usepackage[T1]{fontenc} 
\usepackage[utf8]{inputenc}
\usepackage[dvipsnames,table]{xcolor}

\usepackage{mwe,bm,verbatim,cprotect,xcolor,xspace}
\usepackage{orcidlink}
\usepackage{csquotes,float, ulem}
\usepackage{bbold}
\usepackage{subcaption}
\usepackage[export]{adjustbox}
\usepackage{multirow}
\usepackage{diagbox}

\usepackage{booktabs} 
\usepackage{array} 
\usepackage{geometry} 
\usepackage{graphicx} 
\usepackage{changepage} 

\usepackage{longtable}

\usepackage{color}

\newcommand{\code}[1]{{\texttt{#1}}}
\renewcommand{\emph}[1]{\textit{#1}}

\title{A Designer’s Guide to Lunar Far-Side Interferometer Array: Power Spectrum Measurement and Cosmological Constraints from the Dark Ages}
\author[a]{Yuewei Wen\orcidlink{0000-0003-2579-7039}}
\author[a]{Bin Yue \orcidlink{0000-0002-7829-1181}}
\author[a]{Yidong Xu\orcidlink{0000-0003-3224-4125}}
\author[a,b]{Furen Deng\orcidlink{0000-0001-8075-0909}}
\author[a,b]{Chen Zhang\orcidlink{0009-0005-1288-8873}}
\author[a,b]{Fengquan Wu\orcidlink{0000-0002-6174-8640}}
\author[a,b]{Xuelei Chen\orcidlink{0000-0001-6475-8863}}
\affiliation[a]{State Key Laboratory of Radio Astronomy and Technology, National Astronomical Observatories of China, CAS, 20A Datun Road, Beijing, China 100101}
\affiliation[b]{School of Astronomy and Space Science, University of Chinese Academy of Sciences, Beijing, China 100049}

\emailAdd{wenyw@bao.ac.cn}
\emailAdd{xuelei@cosmology.bao.ac.cn}

\abstract{The 21-cm emission line from neutral hydrogen during the cosmic Dark Ages can be a powerful probe of cosmological models and early universe physics. This work provides a quantitative forecast for the design requirements of a lunar far-side interferometer array aimed at measuring the 21-cm power spectrum and constraining inflationary models through the running of the spectral index $\alpha_s$. During the Dark Ages, larger collapsed objects have not yet formed, allowing linear perturbation theory to remain valid down to much smaller scales than is possible in current large‑scale structure or CMB surveys. We first validate this linearity assumption by quantifying the contribution of minihalos to the 21-cm signal. We then establish a generalized and flexible analytical framework for the baseline density distribution of interferometers that may consist of an arbitrary number of stations or sub-arrays. Incorporating a realistic noise model, we determine the configurations necessary to reach the detection threshold and demonstrate that distributing the total collecting area into multiple stations can improve the signal-to-noise ratio of the power spectrum at a tunable small scale of interest by up to two orders of magnitude. We then show that a lunar array requires at least $\sim30,000$ probed Fourier modes to achieve a constraint on inflation of $\sigma(\alpha_s) = 0.034$, a result competitive with the Planck 2018 results and capable of distinguishing between different inflationary scenarios. We quantitatively explain how thermal noise severely erodes modes at high redshifts and small scales—scales previously considered easily accessible to Dark Ages observations in the literature—and discuss the prospects for Dark Ages observations as a new and independent probe despite this limitation.}


\begin{document}

\date{\today}

\maketitle
\flushbottom
\clearpage

\section{Introduction}

The cosmic Dark Ages are a pivotal yet largely unexplored epoch in the evolutionary history of our Universe. This period begins after the recombination of hydrogen ($z \sim 1100$) and ends before the formation of the first stars (roughly $z \sim 30$)\cite{Barkana:2000fd}. Before gas clouds are reionized during star formation, the baryonic content of the Universe is predominantly neutral hydrogen (HI). Through interactions between HI and photons from the cosmic microwave background (CMB), a spin-flip transition due to the hyperfine structure splitting in hydrogen can either emit or absorb a photon of wavelength 21.1 cm\cite{Pritchard:2008da,Pritchard_2012}. In the absence of any luminous sources, this ``21-cm'' signal becomes almost the sole observable and a most critical tracer of matter density distribution and structural growth during the Dark Ages\cite{Silk:2020bsr}. 

Informing us about physical processes during the first billion years after the Big Bang, this signal holds great potential for addressing fundamental questions in cosmology, concerning inflation, CMB spectral distortions, the physical nature of dark matter and the formation history of supermassive black holes\cite{Barkana:2004zy, Silk:2020bsr, Carr:2018rid, Latif:2016qau, Bernal:2017nec}. Unlike CMB surveys, which are limited to $k \lesssim 0.1$ $h$ Mpc$^{-1}$ by Silk damping and instrumental resolution, or large-scale structure (LSS) surveys, which are confined to $k \sim 1$ $h$ Mpc$^{-1}$ by nonlinear structure formation in the late Universe \cite{DES:2021bvc}, the Dark Ages 21-cm signal can, in principle, access a much larger Fourier volume. As a result, the 21-cm signal from the cosmic Dark Ages holds great potential to become a new and independent probe that not only complements observations from the CMB and late-Universe LSS surveys but also reaches regions previously inaccessible in observational space.

During this epoch, structural growth remains largely linear. Complicated astrophysical processes associated with star formation have not yet begun. Nonlinearity manifests mainly through a sparse population of minihalos (halos too small to cool and form stars) with masses $10^4$–$10^8$ $M_\odot$ \cite{Iliev:2002gj,Furlanetto:2006xi}. Therefore, primordial density fluctuations generated by inflation are well preserved, making the 21-cm signal an excellent probe of cosmic initial conditions. With sufficient angular resolution, one can approach scales on the order of $\sim 1000$ Mpc$^{-1}$. Moreover, in a clean and ideal radio environment, a three-dimensional tomographic survey can cover a wide range of frequency bands, which translates into a large volume in redshift space. In the limit of cosmic variance, one could access $\sim 10^{12}$ independent Fourier modes using 21-cm measurements from the Dark Ages, far exceeding the $\sim 10^8$ modes from Stage-IV LSS surveys and the $\sim 10^6$ modes from the Planck 2018 analysis \cite{Silk:2020bsr, Loeb:2003ya, Planck:2018jri}. 

The 21-cm signal is present only when the spin temperature of the neutral hydrogen gas deviates from that of the CMB. Prior to the formation of the first stars and the onset of radiative processes, this deviation occurs after baryons decouple from the CMB and cool adiabatically as the Universe expands, giving rise to an absorption signal that roughly corresponds to $30 \lesssim z \lesssim 200$. The 21-cm signal in this epoch is redshifted to wavelengths of about 6.5 to 47 meters, or frequencies of 45 to 6 MHz, which correspond to the longest wavelengths of radio astronomical observation. Observations in this low-frequency band have been extremely limited to date because any ground-based radio telescope must deal with absorption and refraction from Earth's ionosphere and eliminate radio frequency interference (RFI) from human activities worldwide. This is precisely why the far side of the Moon—no ionosphere and shielded from RFI from Earth—is an ideal site for observing signals from the Dark Ages.

Proposals to deploy radio telescopes to the far side of the Moon emerged as early as the 1960s \cite{Gorgolewski:1966}. Six decades later, remarkable advancements in space technology have made the construction of lunar surface or cislunar telescopes a conceivable prospect \cite{Chen2019,Koopmans2021,Silk:2020bsr}. China's Chang'e-4 mission made the first landing on the far side of the Moon in 2024 and carried out low-frequency radio experiments from the lunar surface\cite{Zhu2021}, from orbit \cite{Zhang2017,Yan2023} and via the relay satellite \cite{NCLE2018}. Further low-frequency radio experiments on the lunar surface include the Radiowave Observations on the Lunar Surface of the photo-Electron Sheath instrument (ROLSES-1) \cite{Hibbard:2025} and the upcoming Lunar Surface Electromagnetics Explorer (LuSEE-night)\cite{bale2023}. Several low-frequency experiments in lunar orbit have also been proposed, including DARE/DAPPER \cite{Burns2012,Burns:2021pkx}, Hongmeng (also known as Discovering the Sky at Longest Wavelengths, the DSL Project)\cite{Chen2019,Chen2021,Chen2023}, CosmoCube\cite{Artuc2024,Zhu2025}, PRATUSH \cite{Rao2023}, and SEAMS \cite{2021AIPC.2335c0005B}. 
There are also proposals to build large radio telescopes on the lunar surface by using a lunar crater as a reflector \cite{Bandyopadhyay_2021} or by constructing an array of radio telescopes linked by interferometry, such as the Dark Ages Lunar Interferometer (DALI) \cite{Lazio_2009}, the Cosmic-dawn Dark-ages EXplorer (CoDEX) \cite{Koopmans2021}, the Astronomical Lunar Observatory (ALO) \cite{2024AAS...24326401K}, the Large-scale Array for Radio Astronomy (LAFAR) \cite{Chen:2024tvn}, and FarSide/FarView \cite{burns2019,Polidan:2024}.

Cosmological studies based on the Dark Ages demand more than just a measurement of the global 21-cm spectrum; they require the spatial fluctuations—a power spectrum accessible only with an interferometric array. Given the considerable technological and engineering challenges involved in constructing such large radio arrays on the lunar surface, reliable and accurate forecasts that incorporate careful theoretical modeling of the 21-cm signal and realistic levels of noise are essential. This work focuses on developing such forecasts, quantifying the ability of various lunar far-side configurations to measure the power spectrum and constrain cosmological parameters, in particular those related to inflation.

While the evolution of matter density perturbations during the Dark Ages is relatively ``simple'', as in it does not involve complicated star formation and feedback, it still includes the growth of structure under the influence of both dark matter and baryons, the thermal evolution of gas and the coupling between photons and baryons. All of these affect the 21-cm signal. To plan large radio arrays on the lunar surface, a carefully scrutinized predictive framework for the 21-cm signal from the Dark Ages is worthwhile, and we present a comprehensive study in this work.

This paper is structured as follows. Section~\ref{sec:power_spectrum} gives an overview of 21-cm physics during the Dark Ages and explains how the power spectrum is computed, including redshift-space distortions (RSD) and the Alcock--Paczy\'{n}ski effect. In Section~\ref{sec:minihalo}, we analyze the contribution from minihalos and determine the scales up to which linear perturbation theory remains valid. We then turn to observational considerations. Section~\ref{sec:noise_power_spectrum} discusses the modeling of thermal noise for an interferometer array. Section~\ref{sec:baseline_density} presents the baseline density distributions for the various array configurations considered in this work. The impact of foregrounds and the choice of basic survey parameters are addressed in Sections~\ref{sec:foreground} and \ref{sec:parameters}, respectively. Building on these, we present the array designs necessary for detecting the 21-cm power spectrum and study the advantages of different configurations in Section~\ref{sec:Pk_results}. Then, in Section~\ref{sec:alpha_s}, we use the running of the spectral index $\alpha_s$ as a case study to demonstrate the constraining power of Dark Ages observations for inflation, identifying the configurations needed to achieve meaningful constraints and comparing our forecasts to existing CMB results. We conclude in Section~\ref{sec:conclusion}.

Alongside this manuscript, we release a public Python code, \texttt{DarkAgesCosmology}\footnote{\url{https://github.com/wen-yuewei/DarkAgesCosmology.git}}, which enables robust and customizable calculations of the 21-cm signal from the cosmic Dark Ages. The code has the flexibility to model interferometer sensitivity for configurations with any number of stations or sub-arrays and to provide forecasts for cosmological and inflationary parameters.

Throughout this work, we assume a $\Lambda$CDM cosmology and adopt the best-fit values from the Planck 2018 results as our fiducial cosmology: $\Omega_b h^2 = 0.02242$, $\Omega_c h^2 = 11933$, $H_0=67.66$, $\ln{\left(10^{10} A_s\right)} = 3.03$, $n_s=0.9655$, and $\tau=0.053$ \cite{Planck:2018vyg}.

\section{Power Spectrum of the 21-cm Signal}\label{sec:power_spectrum}
In radio astronomy, the intensity of radiation is usually expressed as a brightness temperature $T_b \equiv \left(\lambda^2 / 2 k_B\right) I_\nu$, defined as the temperature of a blackbody that emits the same intensity at the same wavelength. For the 21-cm signal from hydrogen's spin-flip transition, the average brightness temperature relative to the CMB temperature $\delta \bar{T}_b$ in the diffuse intergalactic medium (IGM) is
\begin{align}
    \delta \Bar{T}_b(z) = \frac{T_s(z) - T_\gamma(z)}{1+z} \left(1 - e^{-\tau}\right),
    \label{eq:brightness_temp}
\end{align}
where $T_\gamma$ is the CMB temperature, $T_s$ is the spin temperature, and $\tau$ is the 21-cm optical depth. Fluctuations in this 21-cm brightness temperature, sourced by matter density, gas temperature and radiative effects, contain rich information about the growth of large-scale structure and the cosmic initial conditions and can serve as a critical test of cosmology. 

In this work, we compute the spatial fluctuations in the 21-cm brightness temperature, specifically its cylindrically averaged power spectrum $P_{21}(k_\perp, k_\parallel,z)$, using the numerical package \code{CAMB}\footnote{\url{https://camb.readthedocs.io}}, which solves the Boltzmann-Einstein equations for the evolution of the photon distribution function due to interactions with neutral hydrogen. As the benchmark tool for the linear-theory 21-cm power spectrum of the Dark Ages, \code{CAMB} incorporates many subtle effects that contribute to the power spectrum, including redshift-space distortions, perturbations in the ionized fraction $x_e$ (minimal during the Dark Ages but not entirely negligible) and perturbations in the 21-cm optical depth. A detailed treatment of these effects was given in \cite{Lewis:2007kz}, and we recapitulate only several key steps in the calculation of the power spectrum here. \code{CAMB} includes the baryon peculiar velocity $v_b$, which is numerically equal to the baryon-CDM relative velocity $v_{\mathrm{bc}}$ because the CDM velocity is set to zero in the synchronous gauge. However, in its calculation of the 21-cm power spectrum, $v_b$ contributes only at linear order through RSD and the Doppler effect. The second-order effective viscous damping arising from $v_{\mathrm{bc}}^2$, which is expected to suppress the power spectrum by about 10--20\% on scales $k \gtrsim 10$ Mpc$^{-1}$, is not included\footnote{Not including this suppression does not affect our results. As we will show, the maximum scales we probe in later sections are $k \sim 0.5$ Mpc$^{-1}$, well below the regime where the effect becomes significant. When we do venture into the range $k \sim 1$--$5$ Mpc$^{-1}$ in Sec.~\ref{sec:power_spectrum_SNR}, this effect contributes only a percent-level correction to the signal-to-noise ratio we discuss. Moreover, our analysis focuses on the relative change in S/N between different array configurations rather than on absolute values.}\cite{Tseliakhovich:10}.

The code \code{CAMB} calculates and outputs the transfer function of an isotropic ``monopole'' source $T_{\rm mono}(k)$ of the 21-cm fluctuations, defined as
\begin{align}
     T_{\rm mono}(k)  = \Delta_s + (r_\tau - 1) (\Delta_{\rm HI} - \Delta_{T_s}).
\end{align}
The first term $\Delta_s \equiv \Delta_{\rm HI} + T_\gamma(\Delta_{T_s} - \Delta_{T_\gamma})/(T_s - T_\gamma)$ incorporates perturbations from fluctuations in baryon density ($\Delta_{\rm HI}$), gas temperature ($\Delta_{T_s}$) and photon temperature ($\Delta_{T_\gamma}$), using the notation $\Delta_x$ to denote $\delta x/ \Bar{x}$. The second term in $T_{\rm mono}(k)$ represents contributions from perturbations to the 21-cm optical depth, with the factor $r_\tau \equiv \tau e^{-\tau} / (1 - e^{-\tau})$.

A key quantity that enters both the calculation of the global brightness temperature $\delta \Bar{T}_b$ and its fluctuations $\Delta_s$ is the spin temperature $T_s$. It is the excitation temperature of the spin-flip transition in neutral hydrogen and effectively describes the relative population of the singlet and triplet states. Because the 21-cm transition is driven by collisional interactions between gas particles and with CMB photons (ignoring all radiative effects during the Dark Ages), the spin temperature can be solved by assuming an equilibrium where the population in the singlet or triplet state is conserved, yielding
\begin{align}
    T_s^{-1} = \frac{T_\gamma^{-1} + x_c T_K^{-1}}{1+x_c},
    \label{eq:spin_temp}
\end{align}
where $x_c$ is the collisional coupling coefficient and $T_K$ is the kinetic temperature of the gas. \code{CAMB} extends this expression further by incorporating effects from small changes in the fraction of ionized species and perturbations to the photon phase space density. We refer the reader to Section IV of \cite{Lewis:2007kz} for a detailed derivation of the full $T_s$ and its perturbations $\Delta_{T_s} - \Delta_{T_\gamma}$.   

\subsection{Redshift-Space Distortion and the Alcock-Paczy\'{n}ski Effect}
\label{sec:RSD-AP}

\begin{figure}
    \centering \includegraphics[width=0.8\linewidth]{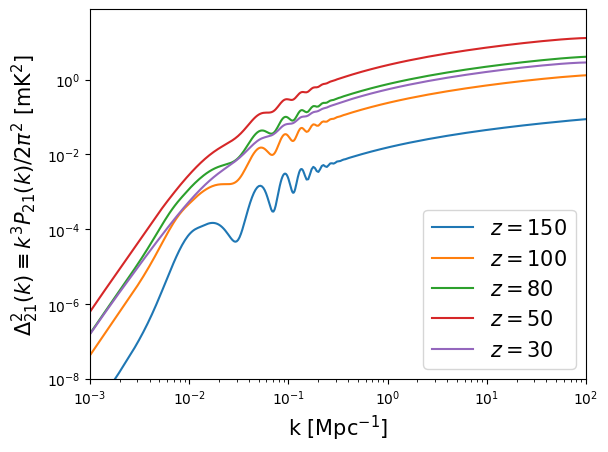}
    \caption{The dimensionless, spherically averaged 21-cm power spectrum at various redshifts during the cosmic Dark Ages calculated using \code{CAMB}.
    } 
    \label{fig:Pk_1D_z}
\end{figure}

In redshift space, perturbations in the brightness temperature also have an anisotropic source from velocity effects, which in Fourier space takes the form $\delta_v(\mathbf{k}) = \mu^2 \delta_b$, where $\mu$ is the cosine of the angle between the wavenumber vector $\mathbf{k}$ and the line-of-sight direction, and $\delta_b$ represents fluctuations in the baryon density. Combining the monopole and the anisotropic sources, fluctuations in the brightness temperature is   
\begin{align}
    \delta_{T_b}(k, \mu) \simeq \delta_{\rm mono}(k) + \mu^2 \delta_b(k).
\end{align}
Defining the 21-cm power spectrum as $(2\pi)^3 \delta^D(\mathbf{k_1} + \mathbf{k_2})P_{21}(\mathbf{k_1}) \equiv \left(\delta \Bar{T}_b\right)^2 \left<\delta_{T_b}(\mathbf{k_1}) \delta_{T_b}(\mathbf{k_2}) \right>$, where $\delta^D$ is the Dirac delta function, it can be expanded as a polynomial in $\mu$:
\begin{align}
    P_{21}(k, \mu) = \left(\delta\bar{T}_b\right)^2 \left[\mu^0 P_{\mu^0}(k) + \mu^2 P_{\mu^2}(k) + \mu^4 P_{\mu^4}(k)\right],
\end{align}
where
\begin{align}
    & P_{\mu^0}(k) = \mathcal{P}_\chi(k) T^2_{\rm mono}(k) \nonumber \\
    & P_{\mu^2}(k) = 2\mathcal{P}_\chi(k) T_{\rm mono}(k) T_b(k) \nonumber \\
    & P_{\mu^4}(k) = \mathcal{P}_\chi(k) T^2_b(k).
\end{align}
In practice, the global brightness temperature $\delta\Bar{T}_b$, the primordial power spectrum $\mathcal{P}_\chi(k)$, the 21-cm monopole transfer function $T_{\rm mono}$ and the baryon transfer function $T_b(k)$ can all be obtained from the output of \code{CAMB}. The spherically-averaged 21-cm power spectrum, integrated over all angles, is 
\begin{equation}
P_{21}(k) = \left(\delta \Bar{T}_b\right)^2 
\left[P_{\mu^0}(k) + (1/3) P_{\mu^2}(k) + (1/5) P_{\mu^4}(k)\right],
\end{equation}
and Fig.~\ref{fig:Pk_1D_z} shows examples of this power spectrum at various redshifts during the Dark Ages. 

In the analysis of the 21-cm signal, it is often necessary to adopt a specific cosmological model, which may differ from the true geometry of the Universe \cite{Alcock_1979}. This potential geometrical distortion scales both in the line-of-sight and transverse directions and will introduce anisotropy, including a $P_{\mu^6}$ moment, into the 21-cm power spectrum. Denoting the power spectrum without this Alcock-Paczy\'{n}ski (AP) effect as $P^{\rm tr}(k)$, the observed power spectrum becomes \cite{Barkana:2005nr}
\begin{equation}
    P_{\rm 21}(k,\mu) = \mu^6 P_{\mu^6}(k,\mu) + \mu^4 P_{\mu^4}(k,\mu) + \mu^2 P_{\mu^2}(k,\mu) + P_{\mu^0}(k,\mu),
    \label{eq:2D_power_spectrum}
\end{equation}
where 
\begin{align}
    P_{\mu^6} & = -\alpha \left(4 P^{\rm tr}_{\mu^4} -\frac{\partial P^{\rm tr}_{\mu^4}}{\partial \log{k}} \right) \nonumber \\
    P_{\mu^4} & = P^{\rm tr}_{\mu^4} + \alpha \left(5 P^{\rm tr}_{\mu^4} - 2 P^{\rm tr}_{\mu^2} + \frac{\partial P^{\rm tr}_{\mu^2}}{\partial \log{k}} \right) - \alpha_\perp \left(3 P^{\rm tr}_{\mu^4} + \frac{\partial P^{\rm tr}_{\mu^4}}{\partial \log{k}} \right) \nonumber \\
    P_{\mu^2} & = P^{\rm tr}_{\mu^2} + \alpha \left(3 P^{\rm tr}_{\mu^2} + \frac{\partial P^{\rm tr}_{\mu^0}}{\partial \log{k}} \right) - \alpha_\perp \left(3 P^{\rm tr}_{\mu^2} + \frac{\partial P^{\rm tr}_{\mu^2}}{\partial \log{k}} \right) \nonumber \\
    P_{\mu^0} & = (1+\alpha) P^{\rm tr}_{\mu^0} - \alpha_\perp \left(3 P^{\rm tr}_{\mu^0} + \frac{\partial P^{\rm tr}_{\mu^0}}{\partial \log{k}} \right),
\end{align}
evaluated at the observed $k$. The AP effect parameters are defined as
\begin{align}
    1 + \alpha & = \frac{H^{\rm tr}(z) D_A^{\rm tr}(z)}{H^{\rm assumed}(z) D_A^{\rm assumed}(z)}, \nonumber \\
    1 + \alpha_\perp & = \frac{D_A^{\rm tr}(z)}{D_A^{\rm assumed}(z)}.
\end{align}

\begin{figure}
    \centering
    \hspace*{-0.8 in}
     \includegraphics[width=1.2\linewidth]{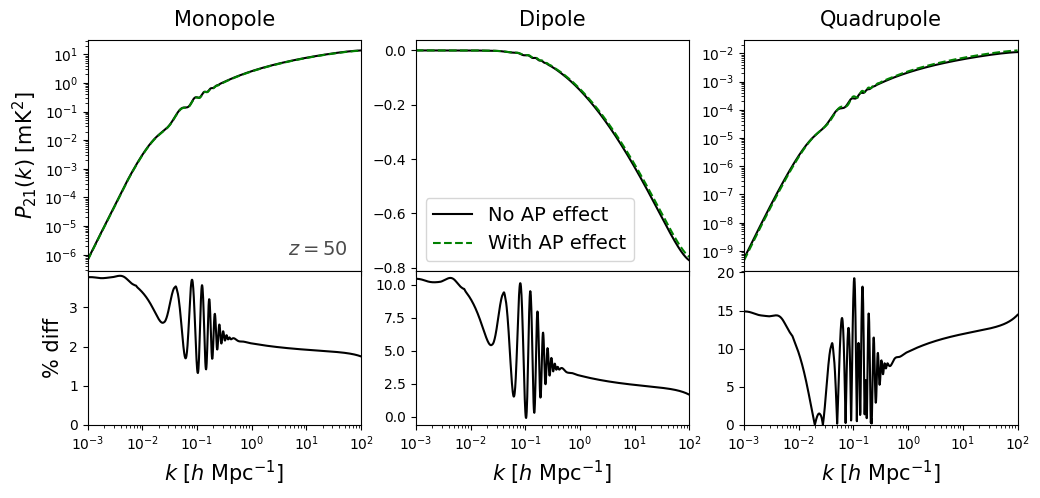}
    \caption{The monopole, dipole, and quadrupole components of the 21-cm power spectrum at $z=50$ with (green) and without (black) the Alcock-Paczy\'{n}ski effect. The bottom panel shows the percent difference in the power spectrum due to the AP effect.}
    \label{fig:AP_effect}
\end{figure}

Fig.~\ref{fig:AP_effect} shows the monopole, dipole and quadrupole parts of the power spectrum at $z=50$ and their percent changes induced by including the AP effect. As expected, the monopole term is the major contributor to the overall 21-cm power spectrum. The AP effect correction is about percent-level in both the monopole and dipole components. Although the AP effect is more pronounced in the third column, the quadrupole term's impact on the total power spectrum is itself subdominant. 

\subsection{The Impact of Minihalos}\label{sec:minihalo}
Even though during the cosmic Dark Ages the perturbations are largely linear and most neutral hydrogen atoms reside in the intergalactic medium (IGM), structure formation is already underway, and a population of minihalos is present, with masses in the range of $10^4$ to $10^8$ $M_\odot$ and radii typically less than $10^{-2}$ kpc. As shown by \cite{Iliev:2002gj} and later substantiated through N-body and hydrodynamic simulations by \cite{Ahn:2005mg}, the intense collisions between gas particles inside minihalos can produce a 21-cm signal distinct from that of the IGM \cite{Yue_2009MNRAS.398.2122Y} and contribute nonlinear effects to the 21-cm power spectrum on small scales. We first quantitatively determine the impact of minihalos and the scales up to which linear perturbation theory remains applicable.

\subsubsection{Minihalo Power Spectrum}

We base our calculations of minihalos on the framework laid out in \cite{Furlanetto:2006xi}, which employs a halo model approach \cite{Cooray:2002dia}. The power spectrum of the 21-cm signal from \emph{only} minihalos consists of a ``one-halo'' term $P_{\rm mh}^{1h}(k,z)$ that accounts for correlations within a single minihalo and a ``two-halo'' term $P_{\rm mh}^{2h}(k,z)$ that accounts for correlations between different halos. These two components at redshift $z$ are given by \cite{Furlanetto:2006xi}
\begin{align}
    P_{\rm mh}^{1h}(k) & = \delta \Bar{T}^2_b \int^{M_{\rm max}}_{M_{\rm min}} dm \left(\frac{m}{\Bar{\rho}} \right)^2 \left[n_h(m) \mathcal{T}_{\rm mh}^2(m) |u(k|m)|^2 \right] \label{eq:minihalo_power_spectrum_1} \\
    P_{\rm mh}^{2h}(k) & = \delta \Bar{T}^2_b P_{\rm lin}(k) I^2_{\rm mh}(k) \label{eq:minihalo_power_spectrum_2} \\
    I_{\rm mh}(k) & = \int^{M_{\rm max}}_{M_{\rm min}} dm \left(\frac{m}{\Bar{\rho}} \right) \left[n_h(m) b(m) \mathcal{T}_{\rm mh}(m) u(k|m) \right],
    \label{eq:minihalo_power_spectrum_3}
\end{align}
where ${\delta T}_b$ is the global 21-cm signal at $z$, $\Bar{\rho}$ is the average matter density of the Universe, $n_h(m)$ is the number density of minihalos, $b(m)$ is the linear halo bias factor, $\mathcal{T}_{\rm mh}^2$ is a temperature factor defined as $(T_s - T_\gamma) / T_s$, and $u(k|m)$ is the Fourier transform of the halo mass profile. Detailed explanations and calculations for most of these quantities are given in Appendix~\ref{app:minihalo}; here we discuss only those that require special clarification. 

For the temperature factor $\mathcal{T}_{\rm mh} \equiv (T_s - T_\gamma) / T_s$ in Eqs.~(\ref{eq:minihalo_power_spectrum_1}), (\ref{eq:minihalo_power_spectrum_2}), and (\ref{eq:minihalo_power_spectrum_3}), the spin temperature $T_s$ is determined by the kinetic gas temperature $T_K$, the CMB temperature, and a collisional coupling constant, as given by Eq.~(\ref{eq:spin_temp}). In \cite{Furlanetto:2006xi}, the gas temperature was set to several values (adiabatic cooling $T_{ad} \equiv 2.58 \left[(1+z)/10\right]^2$ K, 20 K, 100 K, and 1000 K) to demonstrate how the minihalo signal varies with $T_K$.

In this work, we adopt a physical modeling of the gas temperature inside minihalos using the ``TIS'' model (nonsingular, truncated isothermal spheres) proposed by \cite{Iliev:2002gj} and assume the gas temperature equals the virial temperature \cite{Barkana:2000fd}:
        \begin{align}
        T_{\rm vir}(m) = 1.98 \times 10^4 \left(\frac{\mu}{0.6}\right) \left(\frac{m}{10^8 \, h^{-1} M_\odot}\right)^{(2/3)} \left[\frac{\Omega_{m,0}}{\Omega_m(z)} \frac{\Delta_c}{18 \pi^2} \right]^{1/3} \left(\frac{1+z}{10}\right) \textrm{K},
        \end{align}
where the mean molecular weight $\mu=1.22$ for the neutral hydrogen gas during the Dark Ages. $\Delta_c = 18 \pi^2 + 82d -39d^2$ is the ratio of the overdensity at collapse to the critical density, with $d=\Omega_m(z)-1$ \cite{Barkana:2000fd}. The gas temperature profile has a significant impact on the 21-cm signal \cite{Pen_1999, Kim_Pen_2009}, but during the Dark Ages, before any star formation or the presence of rapid radiative cooling from metals, the gas temperature model is simpler than at later epochs. 

In the absence of any ionization or Lyman photon scattering effects, the spin temperature is determined by collisions in the gas and scattering with CMB photons, as described in Eq.~\ref{eq:spin_temp}. To determine the collisional coupling constant $x_c$, we need only consider the dominant collisional processes---those between hydrogen atoms. Thus,
    \begin{align}
    x_c \approx x_c^{HH} = \frac{T_*}{A_{10} T_\gamma} \kappa^{HH}_{1-0}(T_K) n_H, 
    \end{align}
where $T_* = 0.068$ K is the excitation temperature and $A_{10} = 2.85 \times 10^{-15}$ s$^{-1}$ is the Einstein A coefficient \cite{Pritchard_2012}, $n_H= \Bar{\rho}_{\rm mh}/m_H = \Delta_c \rho_{\rm crit}(z)/m_H$ is the number density of hydrogen atoms inside the halo, and $m_H$ is the atomic mass of hydrogen. For the scattering rate $\kappa^{HH}_{1-0}$, when $T_k < 300$ K, we follow the tabulated values provided in Table 2 of \cite{Zygelman_2005}; when $T_k > 300$ K, we employ the fitting formula given by \cite{Zygelman_2005}.

The number density of minihalos, $n_h(M,z)$, is calculated from a halo mass function via $n_h(M,z) = \left(dn/d\ln{M}\right)/M$, and the comoving number density of halos in an interval $dM$ around mass $M$ is given by the Press-Schechter formalism:\cite{Press:1973iz}
\begin{align}
    \frac{dn(M,z)}{d \ln{M}} = \frac{\rho_{m,0}}{M} \left|\frac{dF(M,z)}{d\ln{M}} \right| = \frac{\rho_{m,0}}{M} \left| \frac{d\ln{\sigma(z)}}{d\ln{M}} \right| f(\nu),
    \label{eq:mass_function_general}
\end{align}
where $\rho_{m,0}$ is the present-day matter density, $F(M)$ is the collapse fraction, and $\nu \equiv \delta_c/\sigma(M,z)$.
\begin{eqnarray}
    F(M,z) &=& \frac{1}{\sqrt{2\pi} \sigma(M,z)} \int^\infty_{\delta_c} e^{-\delta^2/2\sigma^2(M,z)} d\delta \\
    f(\nu) &=& \sqrt{\frac{2}{\pi}} \nu e^{\nu^2/2}
\end{eqnarray}
Here, $\sigma(M,z)$ is extrapolated from the present using the linear growth function: $\sigma(M,z) = \sigma(M,0)D(z)$. The rms matter density fluctuation smoothed over a spherical region of radius $R$ enclosing a total mass $M$ today is
\begin{align}
    \sigma^2(M,0) = \int^{\infty}_{k=0} \Delta^2(k) \left(\frac{3j_1(kR)}{kR} \right)^2 d\ln{k},
\end{align}
where $j_1(x) = \sin{x}/x^2 - \cos{x}/x$ is the first spherical Bessel function and $\Delta^2(k) \equiv k^3P(k)/2\pi^2$ is the dimensionless matter power spectrum measured today. The linear growth function is calculated through the growth suppression factor defined as $D(a) \equiv ag(a)/g(a=1)$, and the growth suppression is evaluated using an analytical fitting formula for the $\Lambda$CDM cosmology:\cite{Huterer_2023}
\begin{align}
    g(z) = \frac{2}{5} \frac{\Omega_m(z)}{\Omega_m^{4/7}(z) - \Omega_\Lambda(z) + \left[1 + \frac{1}{2} \Omega_m(z)\right] \left[1 + \frac{1}{70} \Omega_\Lambda(z)\right]},
\end{align}
where $\Omega_m(z)$ and $\Omega_\Lambda(z)$ are the density parameters of matter and dark energy, respectively. 

Alternatively, one may adopt the Sheth–Tormen mass function based on an ellipsoidal collapse model, which predicts stronger fluctuations in the 21-cm brightness temperature \cite{Kim_Pen_2009}. 
\begin{align}
    f(\nu) = A \sqrt{\frac{2q}{\pi}} \left[1 + q \nu^2\right]^{-p} \exp{\left(- q \nu^2/2 \right)}.
\end{align}
For the Sheth-Tormen mass function, we choose the following parameter values: $A = 0.3222$, $q=0.707$, and $p = 0.3$ \cite{Cooray:2002dia}.

\begin{figure}
    \centering
    \includegraphics[width=0.9\linewidth]{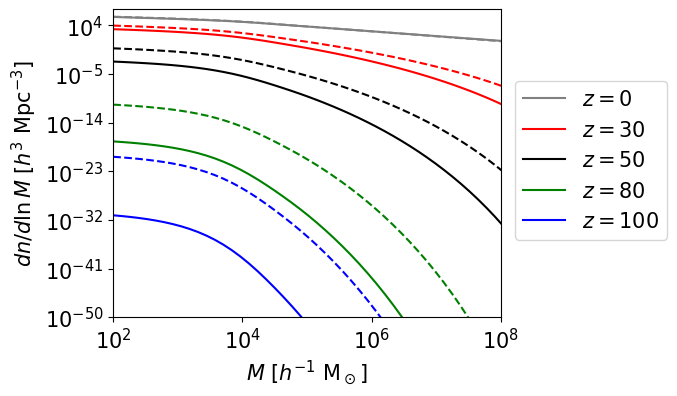}
    \caption{Halo mass function calculated at various redshifts during the cosmic Dark Ages assuming either the Press-Schechter model (solid) or the Sheth-Tormen model (dashed).
    }
    \label{fig:PS_vs_ST}
\end{figure}

In Fig.~\ref{fig:PS_vs_ST}, we show the Press-Schechter and Sheth-Tormen halo mass functions at selected redshifts from $30 \leq z \leq 200$, a period during the Dark Ages that has potential to be observed by radio telescopes. In both cases, the mass function increases substantially as redshift decreases, but the Sheth-Tormen mass function consistently predicts a larger number density of minihalos than the Press–Schechter mass function; at high redshift this difference can be orders of magnitude.

\subsubsection{Comparison with the Total 21-cm Power Spectrum}

Having established the formalism for computing the minihalo contribution to the 21-cm power spectrum, we now ask: how does it compare to the total 21-cm signal from neutral hydrogen atoms in all environments?

To see the full effect of minihalos on very small scales, we extended our calculations to the Jeans scale, below which baryonic gas cannot collapse into halos. The Jeans scale is defined as $k_J \equiv \sqrt{4 \pi G \rho_m(z)} / c_s(z)$, where the sound speed can be estimated by assuming an ideal gas: $c_s(z) = \sqrt{\gamma k_B T_{\rm gas}(z) / \mu m_p}$. During the observable window of the Dark Ages, neutral hydrogen gas in the IGM has already decoupled in temperature from the CMB and begun cooling adiabatically. With $\gamma = 5/3$ and $T_{\rm gas} \sim (1+z)^2$, we estimate the Jeans scale at $z=30$, 50, and 80 to be $k_J \simeq $1500, 1200, and 950 $h$ Mpc$^{-1}$, respectively. Therefore, we choose to calculate the power spectra up to $k = 1000$ $h$ Mpc$^{-1}$ in the comparison that follows.

\begin{figure}
    \centering
    \includegraphics[width=0.9\linewidth]{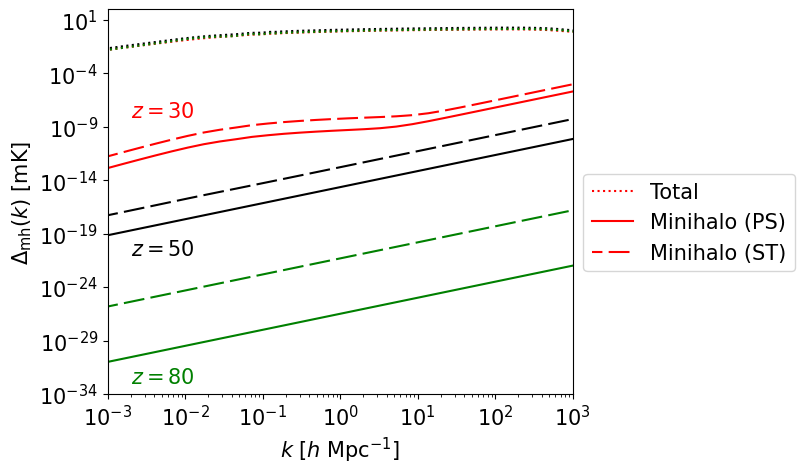}
    \caption{The rms 21-cm brightness temperature fluctuations \emph{solely} from HI atoms in minihalos, assuming a Press-Schechter (solid) or Sheth-Tormen (dashed) mass function, compared with the signal from neutral hydrogen atoms (dotted) in all environments at various redshifts during the Dark Ages. (Owing to the scale of this plot and the faintness of the minihalo signals, the three dotted lines for $z=30$, 50, and 80 appear to overlap; we refer readers to Fig.~\ref{fig:Pk_1D_z} for a more accurate representation.)}
    \label{fig:minihalo_power_spectrum}
\end{figure}

In Fig.~\ref{fig:minihalo_power_spectrum}, we juxtapose the 21-cm signal from minihalos and that from all neutral hydrogen atoms at three different redshifts during the Dark Ages. For minihalos, we show results assuming either the Press-Schechter or the Sheth-Tormen mass function. We plot the rms 21-cm brightness temperature fluctuations $\Delta(k)$ in units of mK (the square root of the power spectrum) to facilitate cross-reference with the results in \cite{Furlanetto:2006xi}.  

Towards the end of the Dark Ages, the number density of minihalos undergoes a relatively rapid increase. At $z=30$, the power spectrum on large scales is dominated by the correlation between different halos, i.e., the two-halo term. Then, a distinct transition occurs around $k \sim 10$ $h$ Mpc$^{-1}$, and the correlation between neutral hydrogen atoms within a single halo begins to dominate. At higher redshifts, minihalos become much more sparse. The impact of the two-halo term thus decreases, and the power spectrum is dominated by the one-halo term. 

At high redshifts during the Dark Ages, the Sheth-Tormen mass function predicts a higher number density of minihalos than the Press-Schechter mass function. Consequently, the 21-cm signal from minihalos computed using the former is stronger. Nevertheless, regardless of the choice of mass function, the brightness temperature fluctuations from minihalos during this epoch are tens of orders of magnitude smaller than the overall signal because the volume fraction of minihalos is too small. Therefore, we conclude that up to the Jeans scale of $\sim 1000$ $h$ Mpc$^{-1}$, the initial clustering of matter in the form of minihalos can be safely neglected in the analysis of the 21-cm brightness temperature power spectrum during the Dark Ages. 

\section{A Lunar Far-Side Interferometer: Instrument and Survey Design}\label{sec:power_spectrum_SNR}

Measuring the 21-cm power spectrum from the cosmic Dark Ages requires an interferometer array. As previously explained, the far side of the Moon offers a unique environment by shielding observations from terrestrial radio interference. In this section, we discuss various plausible array configurations and establish the associated noise model.  

\subsection{Noise Power Spectrum}\label{sec:noise_power_spectrum}
We present the theoretical framework for computing the uncertainty in a 21-cm power spectrum measurement using an interferometer array \cite{Xu:2014bya, Bull:2024eiq}.

For two arbitrary antennas in an interferometer array forming a baseline $\Vec{u}$, the thermal noise when measuring the sky brightness temperature over an observed full bandwidth $\Delta \nu$ is
\begin{align}
    \Delta T_N(\Vec{u}_\perp, u_\parallel) = \frac{\lambda^2(z) \Delta \nu}{A_e} \frac{T_{\rm sys}}{\sqrt{\Delta \nu t_{\Vec{u}}}},
\end{align}
where $\lambda(z)$ is the redshifted wavelength of the 21-cm transition, $A_e$ is the effective collecting area of the antenna, $T_{\rm sys}$ is the system temperature and $u_\parallel$ and $\vec{u}_\perp$ are the line-of-sight and transverse components of this baseline, respectively. The integration time for this specific baseline is related to the $uv$-space resolution and the number of baselines $n_b(\vec{u})$ that sample this particular $\vec{u}$ through $t_{\vec{u}} = (A_e/\lambda^2(z)) n_b(u) t_{\rm int}$, where the integration time per grid is $t_{\rm int} = (\Omega_{\rm FoV}/S_{\rm area}) t_{\rm tot}$, assuming that the survey area $S_{\rm area}$ is larger than the field of view $\Omega_{\rm FoV}$. Thus, for two baselines $\Vec{u}_i$ and $\Vec{u}_j$, their covariance is
\begin{align}
    C^{N}(\Vec{u}_i, \Vec{u}_j) = \left<\Delta T_N(\Vec{u}_i) \Delta T_N(\Vec{u}_j) \right> = \delta_{ij} T^2_{\rm sys} \Delta \nu \left(\frac{\lambda^2(z)}{A_e}\right)^3 \frac{1}{n_b(u_\perp) t_{\rm tot}} \frac{S_{\rm area}}{\Omega_{\rm FoV}}.
\end{align} 

Apart from instrumental thermal noise, the other source of noise is sample variance $C^{\rm SV}(\Vec{u}_i, \Vec{u}_j) = \left<\delta V(\vec{u}_i) \delta V(\vec{u}_j) \right>$. The brightness temperature is related to the visibility through $\delta V(\vec{u}_i) = \int d^3\vec{u} ~ R(\vec{u}_i - \vec{u}) \delta T_b(\vec{u}_i)$, where $R(\vec{u})$ represents a smearing in the $uv$-plane from the main beam pattern. Then, the sample variance can be written as 
\begin{align}
    C^{\rm SV}(\Vec{u}_i, \Vec{u}_j) & = \delta_{ij} \int d^3 \vec{u} ~|R(\vec{u}_i) - \vec{u}| |R(\vec{u}_j) - \vec{u}| \left<\delta T_b(\vec{u}_i) \delta T_b(\vec{u}_j)\right> \nonumber \\
    & = \int d^3 \vec{u} ~|R(\vec{u}_i) - \vec{u}|^2 P_{21}(\vec{u}_i).
\end{align}
Assuming $P_{21}(\vec{k})$ varies on scales much larger than $|R(\vec{u})|^2$, the variance becomes a total integral whose angular part is $\Omega_A = \lambda^2(z)/A_e$ and line-of-sight part is $\Delta \nu$. After converting $P_{21}(\vec{u}_i)$ to $P_{21}(\vec{k}_i)$ using the following change-of-variable relations
\begin{align}
    \vec{k}_\perp = \frac{2 \pi \vec{u}_\perp}{r(z)}, \,\,\,\,\, k_\parallel = \frac{2 \pi H(z)}{\lambda(z)(1+z)} u_\parallel = \frac{2 \pi u_\parallel}{y(z)},
    \label{eq:k_to_u}
\end{align}
the sample variance is
\begin{align}
    C^{SV}(\vec{u}_i, \vec{u}_j) = \frac{\lambda^2(z) \Delta \nu}{A_e} \frac{1}{r^2(z)y(z)} P_{21}(\vec{k}_\perp, k_\parallel),
\end{align}
where $y(z) \equiv \lambda(z)(1+z)/H(z) = \Delta r / \Delta \nu$, and $r(z)$ is the comoving angular diameter distance (please confirm).

With the total covariance matrix $C = C^N + C^{SV}$, we use the Fisher matrix formalism to predict the uncertainty in the 21-cm power spectrum via
\begin{align}
    F_{ij} = \textrm{Tr} \left[ C^{-1} \frac{\partial C}{\partial P_{21}(\vec{k}_i)} C^{-1} \frac{\partial C}{\partial P_{21}(\vec{k}_j)}\right].
\end{align}
We assume that different $\vec{k}$ modes are independent, so only the diagonal terms of the Fisher matrix are nonzero. In this way, the forecast error in the 21-cm power spectrum at wavenumber $\Vec{k}$ is 
\begin{align}
    \delta P_{21}(\vec{k}) = \frac{1}{\sqrt{N (\vec{k})}} \frac{A_e}{\lambda^2(z) \Delta \nu} r^2(z) y(z) \left[C^N(\vec{k}, \vec{k}) + C^{SV}(\vec{k}, \vec{k})\right].
\end{align}
The total number of $k$-modes in a small cylindrical volume $d^3k$ around $\vec{k}$ is $N(\vec{k}) =  k_\perp dk_\perp dk_\parallel V_{\rm survey}/(2 \pi)^2$, where the total survey volume $V_{\rm survey} = S_{\rm area} r^2(z) y(z) \Delta \nu$. 

On this basis, we compute the total measurement error of a cylindrically averaged 21-cm power spectrum in terms of a thermal noise power spectrum $P_N(\vec{k})$:
\begin{align}
    \delta P_{21}(k_\perp, k_\parallel, z) = \frac{2 \pi}{\sqrt{k_\perp dk_\perp dk_\parallel V_{\rm survey}}} \left[P_{21}(k_\perp, k_\parallel, z) + P_N(k_\perp, z)\right],
    \label{eq:total_noise}
\end{align}
where
\begin{align}
    P_N(k_\perp) = T^2_{\rm sys}(z) r^2(z) y(z) \left(\frac{\lambda^2(z)}{A_e}\right)^2 \frac{1}{n_b(u_\perp) t_{\rm tot}} \frac{S_{\rm area}}{\Omega_{\rm FoV}}.
    \label{eq:noise_power_spectrum}
\end{align}

\subsection{Baseline Distribution}\label{sec:baseline_density}
The noise power spectrum of an interferometer (Eq.~\ref{eq:noise_power_spectrum}) depends on the baseline distribution $n_b(k_\perp)$. We first express the density of baselines in the physical plane of the interferometer as a function of physical baseline length $D$, denoted $n_b^{\rm phys}(D)$, with units of m$^{-2}$. This is then converted to the dimensionless $uv$-plane density via $n_b(u) = \lambda^2(z) \, n_b^{\rm phys}(D = u \lambda(z))$ \cite{CosmicVisions21cm:2018rfq}, and finally to a function of $k_\perp$ using Eq.~\ref{eq:k_to_u}.

\paragraph{Random Circular Array} We consider an array with antennas randomly distributed within a circle. For an array of diameter $D_{\rm max}$ containing $N_{\rm ant}$ antennas, the number density of baselines of length $d$ is
\begin{align}
    n_b(d) = \frac{N_{\rm ant}(N_{\rm ant}-1)}{2} \frac{8d}{\pi^2 D^2_{\rm max}} \left[\arccos{\left(\frac{d}{D_{\rm max}}\right)} - \frac{d}{D_{\rm max}} \sqrt{1 - \left(\frac{d}{D_{\rm max}} \right)^2}\right].
    \label{eq:nb_circular}
\end{align}
A detailed derivation is provided in Appendix~\ref{app:circular}. Figure~\ref{fig:baseline_showcase} plots the baseline density distribution for this configuration. 

\paragraph{Approximate Model}
A fitting formula for the baseline density distribution was proposed in \cite{Bull:2024eiq}. The number density of baselines of physical length $d$ is given by
\begin{align}
    n_b^{\rm phys}(d, \{D_0, w\}) = \mathcal{A} (d - D_0)^2 \exp\left[-\left(\frac{d - D_0}{w}\right)^2 \right],
    \label{eq:baseline_FarView}
\end{align}
where the normalization constant $\mathcal{A}$ is fixed by the condition
\begin{align}
    \int_{D_{\rm min}}^{D_{\rm max}} 2 \pi D \, n_b^{\rm phys}(D) \, dD = \frac{N_{\rm ant}(N_{\rm ant} - 1)}{2},
\end{align}
with $N_{\rm ant}$ the total number of antennas. The two parameters $D_0$ and $w$ control the overall shape of the distribution and are chosen by the user. For example, in the Stage-III design proposed by the FarView/FarSide project, these two parameters are set to $D_0 = -2100$ m and $w = 6200$ m. Fig.~\ref{fig:baseline_showcase} also shows several baseline density distributions for different combinations of $(D_0, w)$ and demonstrates their control over the overall shape. As $D_0$ becomes more negative, the distribution skews more towards shorter baselines. Meanwhile, increasing $w$ increases the density of longer baselines. These models have a generally similar shape to the random distribution, though they tend to predict more long baselines.

\begin{figure}
    \centering
    \includegraphics[width=0.9\linewidth]{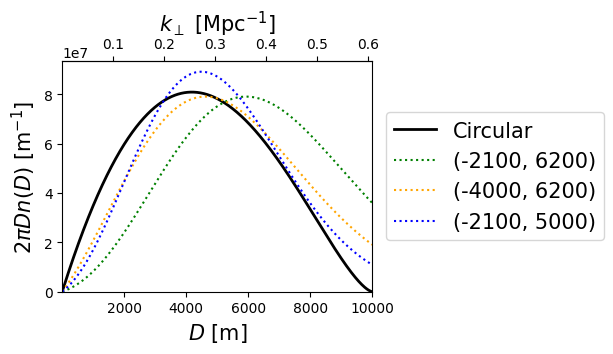}
    \caption{Example configurations of baseline distribution of a circular array with randomly distributed antennas (solid) and of ones following the fitting function adopted by the FarView/FarSide project (dotted) for different combinations of $(D_0, w)$. The top axis indicates the Fourier scale $k_\perp$ corresponding to the physical length of a baseline, evaluated at redshift $z=40$. In each configuration, the array consists of $10^5$ antennas. The minimum and maximum baseline lengths are set to 10 m and 10,000 m, respectively.}
    \label{fig:baseline_showcase}
\end{figure}

\paragraph{Configurations with Two, Three, or More Circular Arrays} A prospective scenario for 21-cm cosmology on the lunar far-side involves multiple stations. It is interesting to explore this possibility and see whether such a configuration can achieve baseline coverage across a range of different scales. Constructing multiple stations may arise from necessity too, for example, if a single suitable lunar site is limited in size. Moreover, it is likely to result from international cooperation, when multiple countries or institutions deploy arrays at different sites on the Moon. The use of multiple distributed sub-stations to achieve flexible baseline coverage and target small angular scales is already a mature technique in ground-based interferometry. For example, LOFAR has 16 remote and 8 international stations extending up to $\sim 1,000$ km from its core array\cite{LOFAR_design}; SKA-Low designed three spiral-arms that extend to form a maximum baseline of 65 km\cite{SKA_low_design}. 

With these prospects in mind and inspired by the designs of ground-based experiments, we examine configurations comprising multiple circular arrays, starting with the simplest case of two identical arrays\footnote{For simplicity, we consider only identical arrays here. This also maximizes the number of long and cross-array baselines, thereby amplifying the advantages of a multi-array configuration over a single-array one.} whose centers are separated by a distance $L$. Each circular array has $N_{\rm ant}$ antennas and a diameter $D_{\rm max}$.

There are two types of baselines: those formed by antennas within a single array (intra-array) and those formed by antennas from two different arrays (cross-array). For intra-array baselines, the density distribution follows the expression from the previous section:
\begin{align}
    n_b^{\rm intra}(d) = 2 \times \frac{N_{\rm ant}(N_{\rm ant}-1)}{2} \times \frac{\mathbb{P}_{\rm intra}(d)}{2 \pi d} = \frac{N_{\rm ant}(N_{\rm ant}-1)}{2 \pi d} \mathbb{P}_{\rm intra}(d),
\end{align}
where the probability distribution function $\mathbb{P}_{\rm intra}(d)$ is the same as in Eq.~\ref{eq:PDF_single_circular}. For cross-array baselines, the density distribution is
\begin{align}
    n_{\rm cross}(d) = N_{\rm ant}^2 \mathbb{P}_{\rm cross}(d;L) = N_{\rm ant}^2 d \int^{2\pi}_0 \frac{\mathbb{P}_{\rm intra}(\sqrt{d^2 + L^2 - 2dL \cos{\phi}})}{2 \pi \sqrt{d^2 + L^2 - 2dL \cos{\phi}}} d\phi,
\end{align}
which can be integrated numerically (derivations provided in Appendix~\ref{app:two_stations}). Summing the two contributions, the total baseline density distribution for two identical arrays separated by $L$ is
\begin{align}
    n_b(d) = n_{\rm intra}(d) + n_{\rm cross}(d).
\end{align}

Building on this categorization, we now give the general expression for the baseline density distribution of an interferometer consisting of $s$ identical stations. Each station is placed at a vertex of an $n$-gon (not necessarily regular) and is itself a circular array of diameter $D_{\rm max}$ containing $N_{\rm ant}$ randomly distributed antennas. The baseline density distribution is:
\begin{align}
    n_b(d) = \frac{1}{2 \pi d} \left[s{N_{\rm ant} \choose 2} \mathbb{P}_{\rm intra}(d) + N_{\rm ant}^2 \sum_{i<j} \mathbb{P}_{\rm cross}(d; L_{ij}) \right],
    \label{eq:nb_generalized}
\end{align}
where $\mathbb{P}_{\rm intra}(d)$ follows Eq.~\ref{eq:PDF_single_circular} and $\mathbb{P}_{\rm cross}$ follows Eq.~\ref{eq:pdf_cross}. The variable $L_{ij}$ denotes the distance between the centers of the $i$-th and $j$-th arrays/stations.

\begin{figure}
    \centering
    \includegraphics[width=\linewidth]{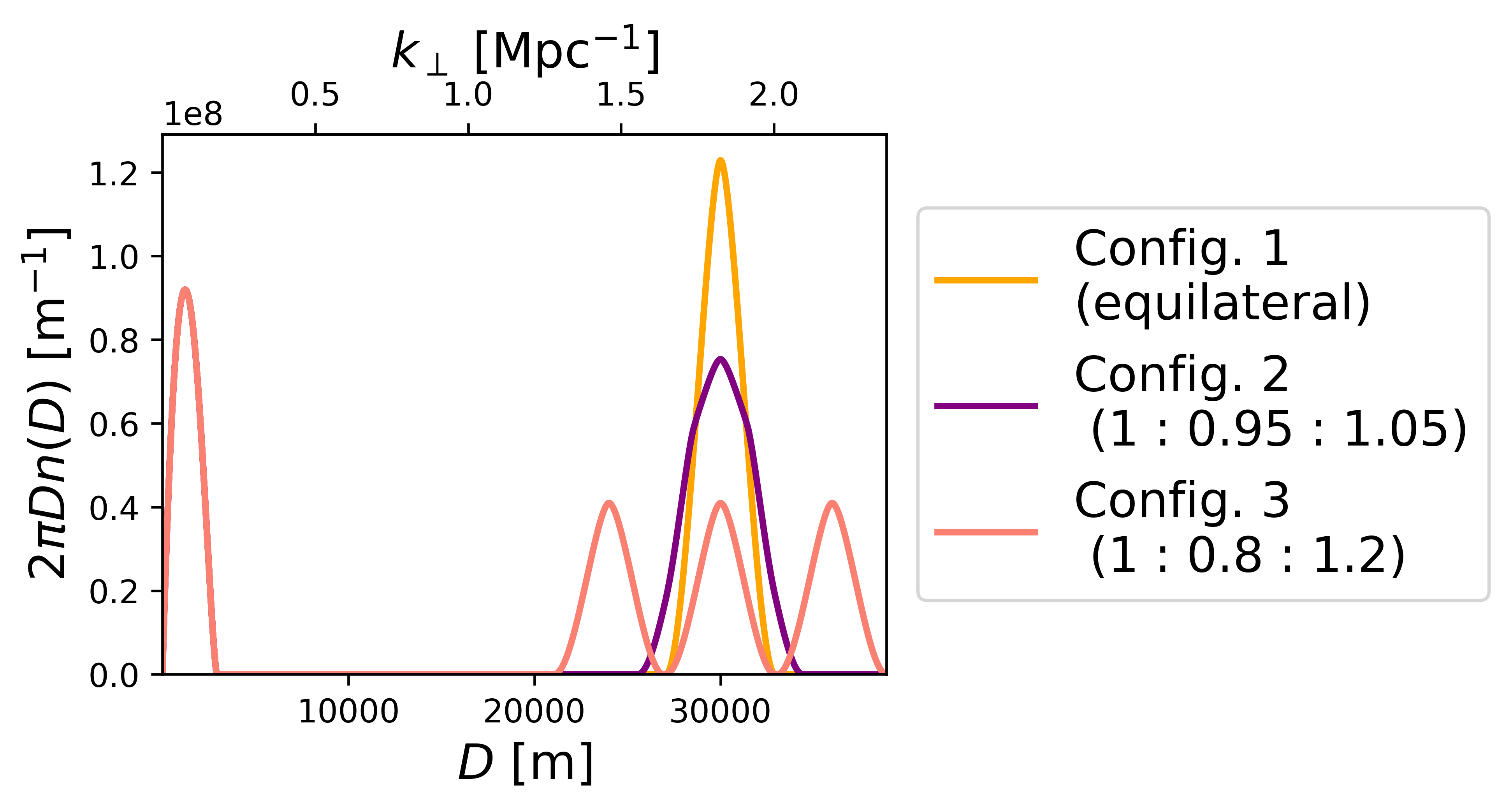}
    \caption{Baseline density distributions for various triple-station configurations, including an equilateral triangle and cases with unequal side lengths. The numbers in parentheses give the ratios of the triangle's side lengths, and the reference side length is $L = 30$ km. The top axis indicates the Fourier scales probed by these baselines at $z = 40$.}
    \label{fig:nb_diff_triangles}
\end{figure}

In a multi-station interferometer, the baseline density distribution develops additional peaks beyond the first one that originates from intra-array baselines. These extra peaks arise from cross-array baselines and their locations are determined by the distances between stations. Fig.~\ref{fig:nb_diff_triangles} illustrates this for a triple-station configuration. When the three stations form an equilateral triangle, all cross-array baselines aggregate into a single narrow peak corresponding to the side length. If the side lengths differ but are close in value, the three sets of cross-baseline peaks formed by each pair of stations will overlap into a broadened feature, as shown in configuration 2. When the side lengths are radically different, each set of cross-baselines forms a distinct, separate peak, as in configuration 3.

\begin{figure}
    \centering
    \includegraphics[width=\linewidth]{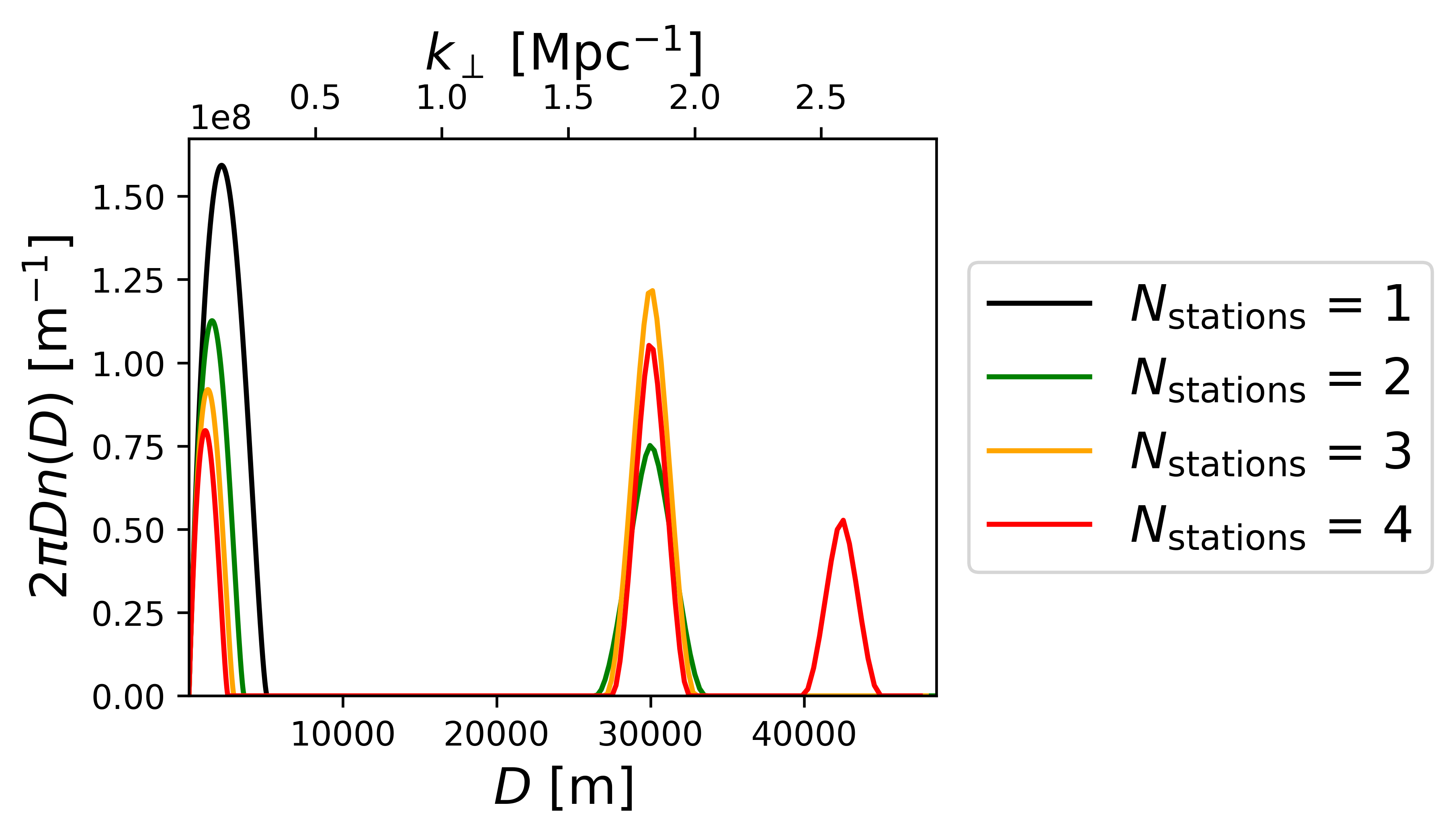}
    \caption{Baseline density distribution for configurations with different numbers of identical circular arrays/stations while keeping the total number of antennas fixed. The side length of each regular $n$-gon is 30 km?. The arrays are all compactly packed with a minimum separation of half a wavelength. The top axis indicates the Fourier scales probed by these baselines at $z = 40$.}
    \label{fig:nb_vs_num_stations}
\end{figure}

Fig.~\ref{fig:nb_vs_num_stations} shows the baseline density distribution for different numbers of stations while keeping the total number of antennas fixed. For visual clarity, all configurations in this plot are assumed to be regular $n$-gons. The triple-station case produces the highest second peak because it maximizes the second term in Eq.~(\ref{eq:nb_generalized}). For $n_{\rm stations} > 3$, the second peak splits into several distinct ones, each lower in amplitude than the single cross-baseline peak of the triple-station configuration. 

\subsection{Foreground Avoidance}\label{sec:foreground}
\begin{figure}[htbp]
    \centering
    \begin{subfigure}[b]{0.49\textwidth}
        \centering
        \includegraphics[width=\textwidth]{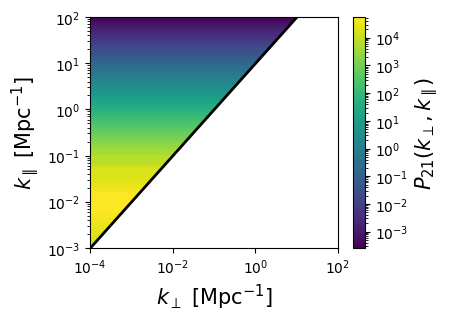}
        \label{fig:foreground_z_40}
    \end{subfigure}
    \begin{subfigure}[b]{0.49\textwidth}
        \centering
        \includegraphics[width=\textwidth]{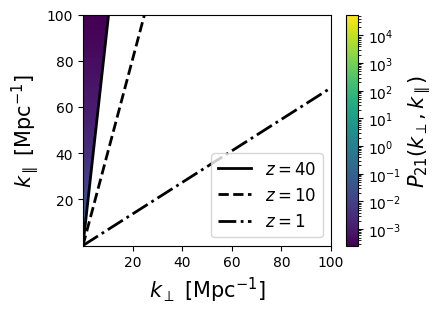}
        \label{fig:foreground_z_linear}
    \end{subfigure}
    \caption{\textit{Left:} Example of the 21-cm power spectrum at $z = 40$ after removing the foreground wedge.
    \textit{Right:} Foreground wedge at $z=1$, 10, and 40, displayed on linear scales in $k_\perp$ and $k_\parallel$ to highlight its evolution.}
    \label{fig:foreground_wedge}
\end{figure}

Interferometers are intrinsically chromatic, and their frequency dependence gives rise to foreground contamination in the 21-cm power spectrum, which takes the shape of a wedge in $k_\perp$-$k_\parallel$ space. The contaminated modes are given by \cite{Obuljen:2017jiy}
\begin{align}
    k_\parallel \leq \frac{H(z) r(z)}{c(1+z)} k_\perp, 
\end{align}
or alternatively in $k$-$\mu$ space where $\mu$ is the cosine of the angle with the line-of-sight direction,
\begin{equation}
    \mu \leq \mu_{\rm min} = \left(\sqrt{1 + \left( \frac{H(z)r(z)}{c(1+z)}\right)^2}\right)^{-1},
\end{equation}
where $H(z)$ is the Hubble parameter and $r(z)$ is the radial comoving distance. Fig.~\ref{fig:foreground_wedge} illustrates the removal of the foreground wedge from the 2D power spectrum at $z = 40$. The wedge occupies the majority of $k$-space, completely excluding information from small scales with $k_\perp \gtrsim 1$. Compounded with restrictions on $k_\parallel$ from the survey's allowed bandwidth and frequency resolution (discussed in the next section), a complete removal of the foreground wedge would substantially reduce the number of Fourier modes available for cosmological analysis. Moreover, as \cite{Pober:2025wui} demonstrates, losing access to modes inside the foreground wedge significantly reduces the detectability of the Dark Ages 21-cm signal by a factor of $\sim 10$. In this work, we take the more optimistic position that future foreground-removal techniques will allow access to all modes up to the interferometer's angular resolution limit. Nevertheless, for completeness, we also include a brief discussion in Sec.~\ref{sec:alpha_s} of results obtained using only $k$-modes outside of the foreground wedge.

\subsection{Interferometer and Survey Specifications}\label{sec:parameters}
We specify the instrumental and survey parameters of our fiducial model in Table~\ref{table:survey_params}. This setup is consistent with the Stage-III design of the FarView/FarSide project \cite{Bull:2024eiq}, which we consider a good balance between ambitious scientific prospects and engineering feasibility in the foreseeable future. 

\begin{table}[ht]
    \caption{Lunar far-side survey and radio array parameters.}
    \begin{center}
        \begin{tabular}{c|c}
          \hline
          \hline
            \textbf{Parameter} & \textbf{Value/Model} \\
         \hline
         Minimum baseline length $D_{\rm min}$ & 4.5 m \\ 
         \hline
         System temperature $T_{\rm sys}$ &  $T_{\rm sys} \simeq T_{\rm sky} = 5000\textrm{K} \, (\nu/50 \textrm{MHz})^{-2.5}$\\
          Number of polarizations $N_{\rm pol}$ & 2 \\
          Gain $G$ & 2 \\
          Effective collecting area $A_e$ & $\lambda^2(z) G / (4 \pi)$ \\
          Field of view $\Omega_{\rm FoV}$ & 10300 deg$^2$ \\
          \hline
          Survey area $S_{\rm area}$ & 20,000 deg$^2$ \\ 
           Fractional bandwidth $\Delta \nu$(\%) & 30\% \\
           Channel width $\delta \nu$ & 10 kHz \\
           \hline
           \hline
        \end{tabular}
    \end{center}  
    \label{table:survey_params}
\end{table}

\section{Power Spectrum Measurement}\label{sec:Pk_results}
Detecting the 21-cm power spectrum from the cosmic Dark Ages is particularly challenging. The signal is intrinsically faint and easily overwhelmed by instrumental noise. Therefore, this section quantifies the signal-to-noise ratios achievable with an interferometric array resembling the one specified in Table~\ref{table:survey_params}. 
The key to overcoming the noise lies in averaging over as many Fourier modes as possible. When the primary goal is detection, this is a worthwhile exchange of angular resolution for detectability. In addition to examining the S/N in individual $k$-bins, we quantify the overall detection significance (in units of $\sigma$) by the Pythagorean sum of signal-to-noise ratios across all bins, following the convention set in \cite{Smith:2025udg}.

\subsection{Detection Significance for a Single Array}

\begin{figure}[htbp]
    \centering
    \begin{subfigure}[b]{0.49\textwidth}
        \centering
        \includegraphics[width=\textwidth]{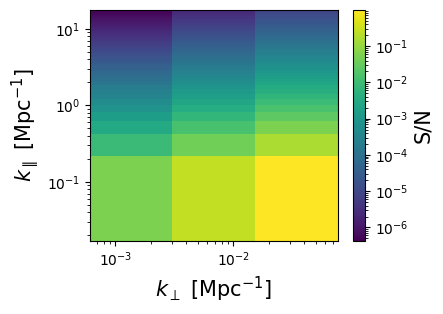}
    \end{subfigure}
    \begin{subfigure}[b]{0.49\textwidth}
        \centering
        \includegraphics[width=\textwidth]{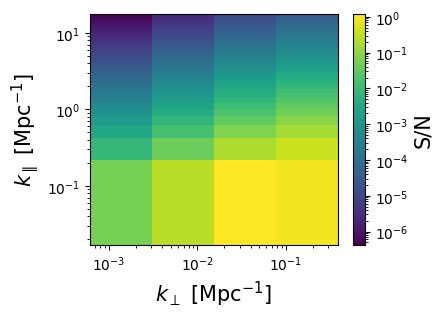}
    \end{subfigure}
    \caption{Signal-to-noise ratios from two array configurations that can reach the detectability threshold of S/N $\gtrsim 1$. \textit{Left}: A compact array of 3.6 km diameter containing $10^5$ antennas, observing for 60,000 hours. 
    \textit{Right}: An array of 10 km diameter containing $2 \times 10^5$ antennas, observing for 100,000 hours. }
    \label{fig:SN_best}
\end{figure}

Figure~\ref{fig:SN_best} presents two single-array designs that can achieve $\mathrm{S/N} \gtrsim 1$ at $z=40$ in key bins and reach $\sim4\sigma$ significance of detection. In both cases, we adopt a transverse bin size $d\log k = 0.7$ Mpc$^{-1}$. 

\begin{itemize}
    \item \textbf{Compact design (left panel):} $10^5$ antennas spaced $\sim10$ m apart (slightly more than half a wavelength at $z=40$), forming an array of diameter $\sim3.6$ km. An integration time of 60,000 hours brings the maximal S/N to $\sim$ 1.
    \item \textbf{Sparse design (right panel):} To access smaller scales ($k_\perp^{\rm max}\sim0.7$ Mpc$^{-1}$), we increase the diameter to $10$ km. This requires $2\times10^5$ antennas and 100,000 hours of integration to achieve the same level of S/N.
\end{itemize} 

This comparison highlights the crucial role of antenna compactness. With a single array, accessing smaller scales or higher $k_\perp^{\rm max}$ requires both a substantial increase in the number of antennas to preserve baseline density and an extension of observation time. Even a one-order-of-magnitude improvement in $k_\perp^{\rm max}$ places a tremendous strain on the required resources.

\subsection{Measuring Small Scales with a Multi-Array Configuration}
Given these resource demands, it is imperative to find a more economical way to measure small scales, which leads us to consider multi-array configurations. They not only align with prospective international collaborations, but also, by sacrificing mid-length baselines, have the ability to generate far more long baselines with the same number of antennas, providing a high signal-to-noise ratio at a \emph{specific, tunable} small scale, mirroring the design philosophy of the ground-based precedents.

\begin{figure}
    \centering
    \includegraphics[width=0.9\linewidth]{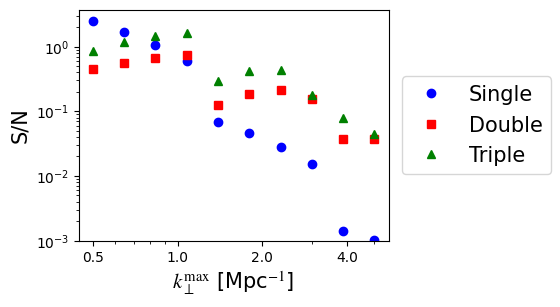}
    \caption{Signal-to-noise ratio at the maximum probed $k_\perp$ scale for single-array (blue), double-array (red), and triple-array (green) configurations. With equal total resources (integration time, number of antennas, etc.), the triple-array configuration consistently outperforms the double array, which in turn outperforms the single array in the regime $k \gtrsim 1$ Mpc$^{-1}$.}
    \label{fig:kmax_SN_array_types}
\end{figure}

Fig.~\ref{fig:kmax_SN_array_types} compares the signal-to-noise ratio at the maximum probed scale for single, double, and triple-array configurations under equal total resources ($3\times10^6$ antennas and 100,000 hours of integration). In the double and triple-array cases, each station is identical and compactly packed (minimum separation set to half a wavelength at $z=40$). For the double array, each of the two stations has $1.5\times10^6$ antennas; for the triple array, each of the three stations has $10^6$ antennas, placed at the vertices of an equilateral triangle.

At $z=40$ and for scales $k \gtrsim 1$ Mpc$^{-1}$, the triple array can improve the S/N of the single array by up to two orders of magnitude. In this regime, the triple array consistently outperforms the double-array configuration, which in turn is better than the single scenario. An equilateral triangle is the only configuration that concentrates all long baselines onto a very narrow range of scales centered on the one corresponding to the side length. With four or more stations, these cross-baselines will split into side lengths and diagonals, diluting the density at the scale of interest, as demonstrated in Fig.~\ref{fig:nb_vs_num_stations}.

\begin{figure}
    \centering
    \includegraphics[width=\linewidth]{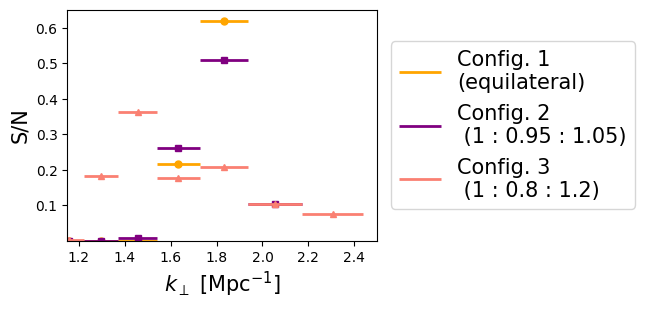}
    \caption{Signal-to-noise ratios at $z=40$ for $k_\perp \gtrsim 1$ Mpc$^{-1}$ bins, assuming various triple-station configurations. The numbers in parentheses indicate the ratios of the triangle's side lengths. The horizontal bars represent the width of each $k_\perp$ bin.}
    \label{fig:SN_triangles_small_scales}
\end{figure}

While the previous figure assumes regular polygons, this need not be the case. Placing stations/sub-arrays at the vertices of non-regular polygons may not only better reflect engineering realities but also offer unique advantages in terms of the flexibility and breadth of coverage in the small-scale regime. Fig.~\ref{fig:SN_triangles_small_scales} takes the triple-station scenario as an example, using the baseline density distributions from Fig.~\ref{fig:nb_diff_triangles} to show the S/N in each $k_\perp$ bin at $k \gtrsim 1$ Mpc$^{-1}$. By adjusting the distances between stations, one can, depending on the specific scientific goal, either sacrifice some maximum S/N for broader small-scale coverage or, by keeping the triangle nearly equilateral, concentrate all baselines into a narrow range to measure fewer $k$-bins with higher significance. This tunability is particularly valuable when the theoretical signal is expected to exhibit features over a range of scales.

Thus, distributing the total collecting area among multiple compact stations is a powerful strategy to achieve high S/N on small scales with great flexibility. This is crucial for testing dark matter models, such as warm or self-interacting dark matter, that predict significant suppression of the power spectrum at $k \gtrsim 1$ Mpc$^{-1}$---a regime not readily accessible to LSS and CMB surveys. A precise measurement in even a single $k$-bin can yield valuable insights.

\section{Constraining Inflation through the Running of the Spectral Index}\label{sec:alpha_s}
We now turn to a potential cosmological application of the measured 21-cm power spectrum from the Dark Ages: constraining inflationary parameters.

Inflation is the leading model for the origin of the cosmos. In this model, primordial perturbations are generated by quantum fluctuations during the inflationary period and become the seeds for subsequent cosmic structure growth. Measurements of these perturbations offer important insight into the extremely early universe, but such data remain limited and are obtained primarily from the CMB. While CMB observations have provided tight constraints on the spectral index $n_s$ and the amplitude of primordial perturbations $A_s$, the running of the spectral index $\alpha_s \equiv dn_s/d\ln k$ remains modestly constrained, with $\sigma(\alpha_s) = 0.012$ from the Planck 2018 analysis \cite{Planck:2018jri}.

The running of the spectral index $\alpha_s$ is a powerful discriminant between different inflationary scenarios. Table~\ref{tab:inflation_models} summarizes the predictions of several models beyond standard slow-roll inflation, illustrating the wide range of $\alpha_s$ values predicted by different theories; we refer the reader to the references listed in the table for more details on each model. Although two of these models predict values of $\alpha_s$ that lie outside the constraints from Planck 2018 analysis, we include them here for completeness. The cosmic Dark Ages 21-cm signal offers a new and independent probe of $\alpha_s$, which can serve as a complementary test of these inflationary models, with the potential to break degeneracies that may arise in current CMB surveys.

\begin{table}[ht]
    \centering    
    \caption{Various beyond-single-scalar-field inflationary theories' predictions of the running of the spectral index $\alpha_s$.}
    \begin{tabular}{c c c}
    \hline
    \hline
    \textbf{Model} & \textbf{Constraints}& \textbf{Reference} \\ 
    \hline
    Scale-invariant $R^2$ inflation & $\alpha_s(k_{\rm pivot}) \sim 10^{-2}$ & \cite{vandeBruck:2021xkm}\\
    \hline
    Multi-natural inflation & \multirow{2}{*}{$-0.06 \lesssim \alpha_s \lesssim$ 0.08} & \multirow{2}{*}{\cite{Czerny:2014wua}} \\
    with sinusoidal functions & & \\
    \hline
    Multi-natural inflation & \multirow{2}{*}{$\alpha_s = -0.013 \pm 0.009$} & \multirow{2}{*}{\cite{Higaki:2015kta}} \\
    with Dedekind eta function & & \\
    \hline
    Linear potential & \multirow{2}{*}{$\alpha_s \sim 0.05$} & \multirow{2}{*}{\cite{Kobayashi:2010pz}}\\
    with periodic oscillations & & \\
    \hline
    Power-law (quadratic) potential & \multirow{2}{*}{$\alpha_s \sim 0.10$} & \multirow{2}{*}{\cite{Kobayashi:2010pz}} \\
    with periodic oscillations &  & \\
    \hline
    Multifield natural inflation & $\alpha_s \sim -5 \times 10^{-3}$ & \cite{Lorenzoni:2024krn} \\
    \hline
    \hline

\end{tabular} \\
\label{tab:inflation_models}

\end{table}

In this section, we assess the ability of a lunar far-side array using 21-cm signals from the cosmic Dark Ages to distinguish between these inflationary models and constrain $\alpha_s$ through the Fisher information matrix formalism. In a single redshift bin, the Fisher matrix is given by 
\begin{equation}
    F_{ij}(z) = \sum_{k_\perp} \sum_{k_\parallel} \frac{\partial P_{21}(k_\perp, k_\parallel, z)}{\partial p_i} \frac{\partial P_{21}(k_\perp, k_\parallel, z)}{\partial p_j} \frac{1}{\delta P_{21}^2(k_\perp, k_\parallel,z)},
\end{equation}
where $p_i$ and $p_j$ denote the parameters of interest while the power spectrum error $\delta P_{21}(k_\perp, k_\parallel, z)$ is given in Eq.~\ref{eq:total_noise}. We then perform a joint analysis of $\alpha_s$ (assuming a fiducial value $\alpha_s = 0$) alongside the relevant cosmological parameters: $\Omega_b h^2$, $\Omega_c h^2$, $H_0$, $A_s$, and $n_s$.

\subsection{Single-Array Constraints}
\begin{figure}
    \centering
    \includegraphics[width=0.7\linewidth]{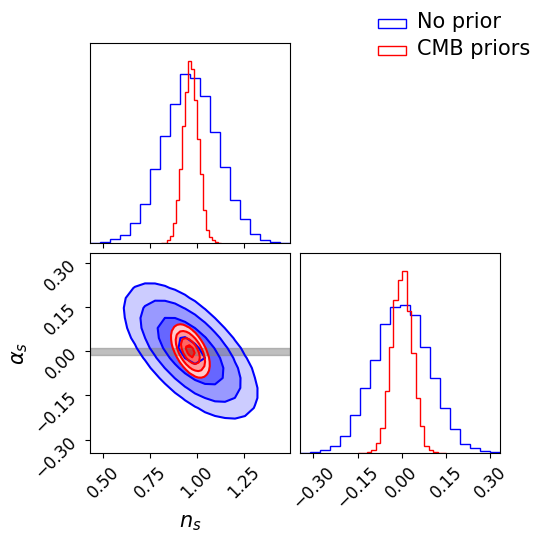}
    \caption{
    Constraints on the spectral index $n_s$ and its running $\alpha_s$ at $68\%$ to $98\%$ confidence levels. Blue contours show the constraints from a single circular array with diameter 5,100 m, $10^6$ randomly distributed antennas, and 100,000 hours of integration time, observing the redshift range $30 \leq z \leq 100$. Red contours include priors from Planck 2018 on the marginalized cosmological parameters $\Omega_b h^2$, $\Omega_c h^2$, $H_0$, and $A_s$. The grey band indicates the Planck 2018 constraint on $\alpha_s$ for reference.}
    \label{fig:ns_alpha_s_single}
\end{figure}

We first consider the single-array scenario. The fiducial array configuration used in the previous section is sufficient for detecting the power spectrum. However, it falls short of what is needed to constrain inflationary models, yielding a weak marginalized constraint of $\sigma(\alpha_s) \sim \mathcal{O}(1)$. Therefore, we now adopt a larger and more capable array comprising $10^6$ antennas as the basis of the discussion in this section. All other aspects of the setup are kept identical to the fiducial design in Table~\ref{table:survey_params}, unless otherwise specified. In Fig.~\ref{fig:ns_alpha_s_single}, we present the joint constraints on $\alpha_s$ and $n_s$\footnote{For a non-zero $\alpha_s$, the spectral index is parameterzied as $n_s(k) = n_s^{(0)} + \alpha_s \ln{(k/k_0)}$. In the discussion that follows, we refer to $n_s$ as the constant term $n_s^{(0)}$ for simplicity.} for this configuration consisting of $10^6$ antennas randomly distributed within a circular array of diameter $5100$ m, with 100,000 hours of integration time and covering the redshift range $30 \leq z \leq 100$. This setup represents the minimal configuration required to achieve a constraint on $\alpha_s$ comparable to that of the latest Planck analysis \cite{Planck:2018jri} and to effectively distinguish between the inflationary models presented in Table~\ref{tab:inflation_models}. 

In Fig.~\ref{fig:ns_alpha_s_single}, we show $1\sigma$ to $3\sigma$ contours from the 21-cm measurement alone alongside those obtained after incorporating Planck 2018 results (TT,TE,EE+lowE+lensing+BAO) on $\Omega_b h^2$, $\Omega_c h^2$, $H_0$, and $A_s$ as priors. Adding priors on the matter-density parameters can break their degeneracy with $\alpha_s$, as they all affect the turnover shape of the matter power spectrum. For the Hubble constant, expansion at high redshift is matter-dominated, so we need external information to pin down $H_0$. Furthermore, since the CMB has already provided precise measurements of the amplitude of primordial fluctuations $A_s$, we incorporate this constraint as a prior to lift degeneracies between inflationary parameters as well. Without the CMB priors, the marginalized constraint on $\alpha_s$ is 0.091 at 68\% CL. Incorporating these priors tightens this constraint to 0.034, an improvement by nearly a factor of two. For completeness, we also report the marginalized constraint on $\alpha_s$ using only modes outside the foreground wedge: $\sigma(\alpha_s) = 0.22$, which underscores the importance of accessing smaller angular scales in the transverse direction. 

Table~\ref{tab:single_array_results} summarizes the constraints on $\alpha_s$ for several other single-array configurations, exploring variations in array size, number of antennas and models of baseline density distribution. The first three entries compare different baseline distribution models and show that the choice of distribution has little impact on $\sigma(\alpha_s)$. In the fourth entry, we increase the array diameter from $5.1$ km to $10$ km while keeping the number of antennas fixed, reducing the antenna density by a factor of $5$; the constraint on $\alpha_s$ worsens by a factor of $\sim2$. The final entry corresponds to the FarView/FarSide Stage-III design. With fewer antennas and a far less compact configuration, this setup yields the weakest constraint.

\begin{table}[ht]
    \caption{
    Marginalized constraints on $\alpha_s$ for various single-array configurations. The other aspects of the array configuration follow those used in Fig.~\ref{fig:ns_alpha_s_single}.
}
    \begin{center}
        \begin{tabular}{c|c|c|c|c}
          \hline
          \hline
           \textbf{Diameter} [m]  & $\mathbf{N_{\rm ant}}$ & \textbf{Density of Ant.} [m$^{-2}$] & \textbf{Baseline Dist. Type} & $\mathbf{\sigma(\alpha_s)}$ \\
         \hline
         5,100 & $10^6$ & 0.05 & Circular & 0.034 \\
         5,100 & $10^6$ & 0.05 & FarView (-1000m, 3100m) & 0.041 \\
         5,100 & $10^6$ & 0.05 & FarView (-1500m, 3100m) & 0.038 \\
         10,000 & $10^6$ & 0.01 & Circular & 0.057 \\
         10,000 & $10^5$ & 0.001 & FarView (-2100m, 6200m) & 9.4 \\
         
           \hline
           \hline
        \end{tabular}
    \end{center}
    
    \label{tab:single_array_results}
\end{table}

\subsection{Multi-Array Constraints}
We now consider configurations of multiple identical circular arrays connected via interferometry---a scenario that fits future international collaboration plans for cosmology from the lunar far side. 

\begin{figure}
    \centering
    \includegraphics[width=0.7\linewidth]{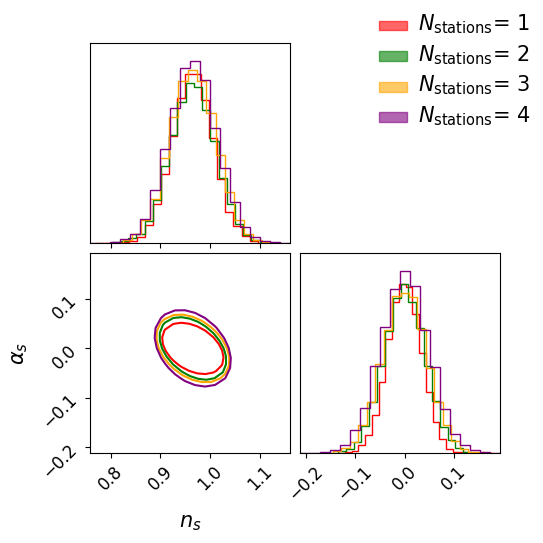}
    \caption{
    $68\%$ confidence contours after incorporating priors from Planck 2018 for the spectral index $n_s$ and its running $\alpha_s$, derived from interferometer configurations consisting of 1 to 4 stations. The total number of antennas is fixed at $10^6$ across all four scenarios; the integration time is 100,000 hours; and the side length of the $n$-gon formed by the arrays is set to 10 km. }
    \label{fig:fisher_multi_array}
\end{figure}

Fig.~\ref{fig:fisher_multi_array} compares the 68\% confidence contours on $n_s$ and $\alpha_s$ (after incorporating Planck 2018 priors) for array configurations with $1$ to $4$ separate stations. For visual clarity, we show only the $68\%$ contours rather than the conventional three nested ones. In all cases, the total number of antennas is fixed at $10^6$, the integration time is 100,000 hours, and each station is placed at the vertices of a regular polygon with side length $10$ km. Separating the available antennas into multiple stations does not have a major impact on the constraint on $\alpha_s$. The constraint progressively worsens as the total collecting area is split into more stations, indicating that it is most advisable to consolidate all available antennas into a single compact array. 

In Table~\ref{tab:multi_array_results}, we present the constraints on $\alpha_s$ for configurations with different numbers of stations and for varying separation distances $L$ between stations. The latter proves to be a very minor factor, though smaller separations yield slightly tighter constraints because such a configuration more closely resembles a single circular array.

\begin{table}[ht!]
    \centering
    \caption{Marginalized constraints $\sigma(\alpha_s)$ (including Planck 2018 priors) for configurations with $1$ to $4$ stations and varying separation $L$. The setup is the same as in Fig.~\ref{fig:fisher_multi_array}.}
    \begin{tabular}{c|cccc}
        \hline
        \hline
        $\mathbf{L}$ [km] & \textbf{1 Station} & \textbf{2 Stations} & \textbf{3 Stations} & \textbf{4 Stations} \\
        \hline
        5  & \multirow{4}{*}{0.034} & 0.039 & 0.041 & 0.044 \\
        10 &  & 0.041 & 0.045 & 0.049 \\
        20 &  & 0.041 & 0.047 & 0.051 \\
        100&  & 0.041 & 0.047 & 0.052 \\
        \hline
        \hline
    \end{tabular}
    \label{tab:multi_array_results}
\end{table}

\subsection{Discussion} 
The 21-cm signal from the Dark Ages has long been considered a probe with huge potential for studying inflation through $\alpha_s$ because, in principle, the immense survey volume at high redshift allows access to up to $\sim 10^{12}$ Fourier modes, far exceeding the $\sim 10^6$ modes used in the latest Planck analysis \cite{Planck:2018jri, Silk:2020bsr}. In practice, however, thermal noise greatly reduces the effective number of accessible modes. Inspired by the Planck estimator $N_{\rm modes}^{\rm XY}(\ell) \equiv 2 \sum_{\ell'=2}^{\ell} \left( C_{\ell'}^{\rm XY} / \delta C_{\ell'}^{\rm XY} \right)^2$ (where XY can be TT, EE, TE, etc.), we define the {\bf effective number of independent Fourier modes} measured in a 21-cm observation weighted by signal-to-noise as
\begin{align}
    N_{\rm modes} = \sum_z \sum_{k_\perp, k_\parallel} \left(\frac{P_{21}(k_\perp, k_\parallel,z)}{\delta P_{21}(k_\perp, k_\parallel,z)}\right)^2.
\end{align}

\begin{figure}
    \centering
    \includegraphics[width=0.9\linewidth]{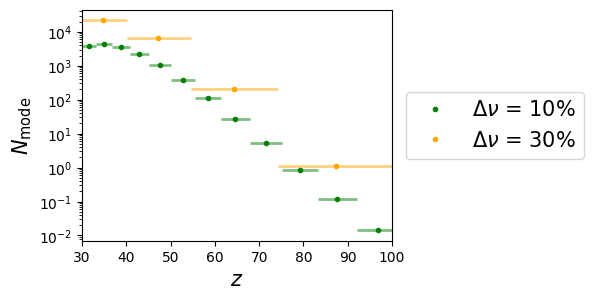}
    \caption{The number of effective Fourier modes probed in each redshift bin in the range $30 < z < 100$, assuming an effective bandwidth of 10\% and 30\%, respectively. The rest of the array configuration parameters are the same as those used in Fig.~\ref{fig:ns_alpha_s_single}.}
    \label{fig:N_modes_vs_z}
\end{figure}

We first quantitatively examine why the large survey volume during the Dark Ages does not automatically translate into a much improved constraint on $\alpha_s$. Using the same setup as the single-array design that produced the contour plot in Fig.~\ref{fig:ns_alpha_s_single}, we trace the number of accessible modes per redshift bin in Fig.~\ref{fig:N_modes_vs_z}, assuming effective bandwidths of either $30\%$ or $10\%$. In both cases, the number of modes drops rapidly from $\sim 10^4$ around redshift 40 to just a few by $z \sim 100$. Thus, the advantage of surveying a wide redshift range quickly diminishes at higher redshifts. The total number of independent modes accessed is $N_{\rm modes} = 1.54 \times 10^4$ for the 10\% bandwidth and $2.83 \times 10^4$ for the 30\% bandwidth. 

\begin{figure}
    \centering
    \includegraphics[width=0.9\linewidth]{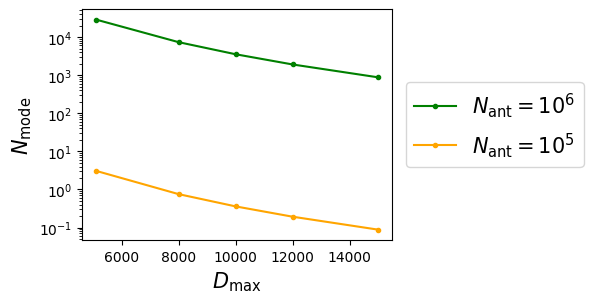}
    \caption{Total number of accessible Fourier modes in the redshift range $30 < z < 100$ plotted against array diameter, with the number of antennas fixed at $10^5$ (orange) and $10^6$ (green). The rest of the array configuration parameters are the same as those used in Fig.~\ref{fig:ns_alpha_s_single}.}
    \label{fig:N_modes_vs_D_max}
\end{figure}

Another advantage of measuring primordial fluctuations during the Dark Ages is that structure growth remains linear down to very small scales, promising more accessible Fourier modes. However, as Fig.~\ref{fig:N_modes_vs_D_max} shows, for a fixed number of antennas, $N_{\rm modes}$ decreases rapidly as we increase the array size to reach smaller angular scales. Increasing the array diameter from 5 to 15 km raises $k_\perp^{\rm max}$ from 0.4 to 1.2 Mpc$^{-1}$ but reduces the total number of accessible modes by a factor of 100. Consequently, when the number of antennas is limited, a smaller but more compact array is more suitable for cosmological constraints despite its inability to access smaller scales.

In summary, the effective number of Fourier modes accessible by a lunar far-side array during the Dark Ages falls far short of the $\sim 10^6$ modes probed by Planck, let alone the often-cited $\sim 10^{12}$ modes. This limitation imposed by thermal noise has been noted in previous works as well. For example, \cite{Smith:2025udg} forecast the detection significance of the 21-cm power spectrum for various array configurations and find a maximum of $26\sigma$ for their best design, corresponding to roughly 700 independent Fourier modes. Perhaps sufficient for a detection, this mode count is far from meeting the needs for competitive constraints on cosmological parameters. Similarly, \cite{Koopmans2021} estimated the signal-to-noise ratio of the power spectrum based on the CoDEX configuration and concluded that high S/N is difficult to achieve. 
Recognizing this challenge, some forecast studies choose to work in the limit where thermal noise is subdominant to cosmic variance. \cite{Bull:2024eiq}, for instance, demonstrated that with the Stage-III FarView/FarSide design, the 21-cm signal would be below the detection threshold, and their forecasts for primordial non-Gaussianity $f_{\rm NL}$ are effectively upper limits under ideal assumptions of no instrumental noise.

Thermal noise is indeed the primary bottleneck in unlocking the full potential of observations in the Dark Ages. The 21-cm line is redshifted to low frequencies where the sky temperature is very high, giving rise to large thermal noise and the necessity for extremely long integration. Even when the underlying survey volume is large, thermal noise erodes the number of Fourier modes when one attempts to reach higher redshift or smaller angular scales. As a result, we find it imperative that this difficulty be carefully quantified, as done in this work, because doing so lays the foundation for mitigating its impact in the future design of lunar missions.

Nevertheless, what is remarkable about this probe is that with only $\sim$ 29,000 effective modes (for the configurations in Fig.~\ref{fig:ns_alpha_s_single} and Fig.~\ref{fig:fisher_multi_array}), the constraint on $\alpha_s$ is already comparable to that of the CMB, implying immense potential yet to be exploited. Although its signal-to-noise ratio remains a major limitation, the Dark Ages signal can access scales that are smaller and larger than the CMB (if not as high $k$ as originally hoped) and is able to trace the full shape of the power spectrum with better capacity. In this sense, it still provides a new and independent lever for constraining primordial fluctuations. Knowing that the number of effective modes decays rapidly at high redshift, one promising future direction is to observe at lower redshifts during the Cosmic Dawn, where thermal noise naturally declines. The trade-off is that the onset of radiation introduces additional physical complexity and small-scale effects such as minihalos will play a more significant role. Despite these challenges, the improved signal-to-noise ratio could bring us closer to the full $\sim 10^{12}$ modes available in the cosmic variance limit.

\section{Conclusion}\label{sec:conclusion}
The 21-cm signal from the cosmic Dark Ages offers a unique possibility to address fundamental questions in cosmology. In this work, we examined the physical processes governing neutral hydrogen during this epoch and assessed the ability of a future lunar far-side array to measure the two-dimensional 21-cm power spectrum and constrain inflationary parameters.

We first quantified the contribution from neutral hydrogen residing in minihalos, using the truncated isothermal spherical model for the kinetic temperature of the gas inside minihalos to calculate its 21-cm emission. Testing both the Press-Schechter and the Sheth-Tormen mass functions, we found that while they predict different halo number densities, the minihalo contribution to the overall 21-cm signal remains negligible up to the Jeans scale ($k_J \sim 1000$ $h$ Mpc$^{-1}$) throughout the observable window of the Dark Ages.

Building on this, we introduced several possible configurations for future lunar far-side arrays and derived a generalized analytical framework for the baseline density distribution of an interferometer comprising an arbitrary number of identical circular stations or sub-arrays. We then forecast their capability to measure the 21-cm power spectrum. We found that for a single array with randomly distributed antennas, observing for 60,000 hours, at least $10^5$  antennas in a circular area of diameter $3.6$ km are required to achieve a signal-to-noise ratio $\gtrsim 1$ at $z=40$ and reach the detection threshold. If one desires to reach smaller angular scales ($\sim 0.7$ Mpc$^{-1}$), a more ambitious and costly design of $2\times10^5$ antennas integrated over 100,000 hours is necessary. As a more economical alternative, a multi-array configuration (e.g., three stations at the vertices of an equilateral triangle) can improve sensitivity at tunable small scales by up to two orders of magnitude compared to a single array. Adjusting station separations adds versatility in small‑scale coverage, which is especially valuable for testing dark matter models that predict suppressed small-scale power.


We then forecast constraints on the running of the spectral index $\alpha_s$ using realistic thermal noise modeling---often optimistically estimated or ignored in previous works. Considering both a single circular array, with baseline distributions following either random antenna distributions or the approximate fitting formula for FarView, and multi-array configurations which may be constructed in the framework of international collaboration of several countries or organizations on the Moon, we also explored how design parameters, including array diameter, number of antennas, integration time, and baseline density distribution, affect the resulting constraints. Our results show that for a single circular array with $10^6$ antennas, a diameter of 5,100 km reaching a maximum angular scale of $k_\perp^{\rm max} \sim 0.4$, and 100,000 hours of integration, the constraint on $\alpha_s$ can reach $\mathcal{O}(-2)$ after surveying the redshift range $30 < z < 100$. After incorporating priors from the Planck 2018 analysis to pin down $H_0$ and break degeneracy with matter density, the constraint becomes $\sigma(\alpha_s) = 0.03$, sufficient to distinguish between standard slow-roll and other exotic inflationary models. 

A key insight from our work is that the accessible Fourier modes from the Dark Ages are limited by the large thermal noise in practice. The effective number of modes measured by the array configurations discussed in this work is only on the order of $10^4$. Thermal noise grows rapidly as one goes to longer baselines or higher redshift, drastically reducing the effective $N_{\rm mode}$. For a fixed number of antennas, expanding the array diameter to reach smaller scales actually reduces the total number of detectable modes, making a compact design preferable for cosmological constraints. Despite the ability of multi-array configurations to greatly improve sensitivity on specific small scales, they cannot tighten the constraints on $\alpha_s$ significantly over a single compact array. In fact, for a given number of total antennas, the constraining power of a single-array design is slightly better because its compactness maximizes the number of Fourier modes available in the analysis.

Despite these limitations, the 21-cm signal from the cosmic Dark Ages remains a powerful probe of the precious information on the primordial Universe. 
We have released the accompanying code, \texttt{DarkAgesCosmology}, which enables flexible testing of cosmological constraints and power spectrum measurement, and we welcome its use by the community.

\section*{Acknowledgments}
This work is supported by China's Space Origins Exploration Program Nos. GJ11010405 and GJ11010401, the National Natural Science Foundation (NSFC) grants (No.12361141814, 12533001, 12421003), the National Key R\&D Program of China No. 2022YFF0504300, and by the Specialized Research Fund for the State Key Laboratory of Radio Astronomy and Technology.

\appendix

\section{Calculating the 21-cm Power Spectrum of Minihalos}\label{app:minihalo}

For quantities in Eqs.~\ref{eq:minihalo_power_spectrum_1} to \ref{eq:minihalo_power_spectrum_3}:
\begin{itemize}
    \item $\delta \Bar{T}_b$: the global 21-cm signal is calculated by running \code{CAMB} twice with the same cosmology to obtain one power spectrum in units of mK$^2$ and one that is dimensionless. The global signal is the square root of the ratio of these two 21-cm power spectra. 
    
    \item $M_{\rm max}$ and $M_{\rm min}$: The minimum mass of minihalos considered in this work is set to the Jeans mass, while the largest minihalo allowed under the TIS model has a virial temperature around $\sim 10^4$ K. We use the empirical formulas provided in \cite{Iliev:2002gj}:
    \begin{align}
        M_{\rm min} & = 5.7 \times 10^3 \left(\frac{\Omega_m h^2}{0.15}\right)^{-1/2} \left(\frac{\Omega_b h^2}{0.02}\right)^{-3/5} \left(\frac{1+z}{10}\right)^{3/2} \, M_\odot \\
        M_{\rm max} & = 3.95 \times 10^7 \left(\frac{\Omega_m h^2}{0.15}\right)^{-1/2} \left(\frac{1+z}{10}\right)^{-3/2} \, M_\odot.
    \end{align}
    
    \item $\Bar{\rho}$ is the mean density of the Universe at redshift $z$: $\Bar{\rho} = \rho_{\rm crit,0} \Omega_{m,0} (1+z)^3$, where $\rho_{\rm crit,0} = 3 H_0^2/8 \pi G$ is the present critical density.
    
    \item $u(k|m)$ is the normalized Fourier transform of the Navarro-Frenk-White (NFW) density profile. In real space, the density profile is \cite{Barkana:2000fd}
    \begin{align}
    \rho(r,z|M) = \rho_{\rm crit}(z) \frac{\delta_c}{c_N x(1+c_Nx)^2},
    \end{align}
    where $x = r/r_{\rm vir}$, the concentration $c_N(M,z) = \left(M/10^9 M_\odot\right)^{-0.1} (25/1+z)$, the characteristic density $\delta_c = (1/3)\Delta_c c_N^3/(\ln{(1+c_N)} - c_N/(1+c_N))$, and the virial radius 
    \begin{align}
        r_{\rm vir} = 0.784 \left(\frac{m}{10^8 \, h^{-1} M_\odot}\right)^{(1/3)} \left[\frac{\Omega_{m,0}}{\Omega_m(z)} \frac{\Delta_c}{18 \pi^2} \right]^{-1/3} \left(\frac{1+z}{10}\right)^{-1} h^{-1} \, \textrm{kpc}.
    \end{align}
    After Fourier transform, \cite{Giocoli_2010}
    \begin{align}
    u(k|M) = \int^{r_{\rm vir}}_0 \frac{4 \pi r^2}{M} \frac{\sin{kr}}{kr} \rho(r|M) dr.
    \end{align}
    
    \item The linear matter power spectrum $P_{\rm lin}(k)$ is computed using \code{CAMB}.
    
    \item The linear halo bias factor \cite{Mo:1995cs}
    \begin{align}
    b(M, z) = 1 + \frac{\nu^2(z) - 1}{\delta(z)},
    \end{align}
    where $\delta(z) = \delta_c D(z)$ and $\nu(z) = \delta(z)/\sigma(M)$.
\end{itemize}

\section{Baseline Density Distribution}
\subsection{Circular Random Array}\label{app:circular}
For a circular array of radius $R$ and diameter $D=2R$ consisting of $N$ randomly distributed antennas, the probability that the distance between two points $A$ and $B$ in the array is less than or equal to an arbitrary distance $d$ is
\begin{align}
    \mathbb{P}(|\vec{A} - \vec{B}| \leq d) & = \iint_s \iint_s \mathbb{P}(A) \mathbb{P}(B) \textbf{1}_{|\vec{A} - \vec{B}| \leq d} dA dB \nonumber \\
    & = \iint_s \iint_s \left(\frac{1}{\pi R^2} \right) \left(\frac{1}{\pi R^2} \right) \textbf{1}_{|\vec{A} - \vec{B}| \leq d} dA dB,
\end{align}
where $\mathbb{P}(A) = \mathbb{P}(B) = 1/\pi R^2$ is the probability density of finding an antenna at position $A$ or $B$. The integral contains a 2D step function $\textbf{1}_{|\vec{A} - \vec{B}| \leq d}$ that returns 1 when the distance between the two points is less than $d$ and 0 otherwise; $dA$ and $dB$ are infinitesimal area elements around $A$ and $B$. 

Fixing point $A$ and considering the inner integral $\iint_s \textbf{1}_{|\vec{A} - \vec{B}| \leq d} dB$, this is equivalent to the intersection area $S(\vec{A}, d) = S(r, d)$ between the array and the small circle of radius $d$ centered at $A$. This area function is isotropic and depends only on $r$, the radial distance from $A$ to the center of the array. Thus, the cumulative probability distribution becomes
\begin{align}
    \mathbb{P}(|\vec{A} - \vec{B}| \leq d) & = \left(\frac{1}{\pi R^2}\right)^2 \iint_s S(r_A, d) dA = \frac{2}{\pi R^4} \int^R_0 r S(r,d) dr.
\end{align}
The area function $S(r, d)$ depends on the relationship between $r$, $d$, and $R$. When $d \leq R-r$, $S(r,d) = \pi d^2$; when $d \geq R+r$, $S(r,d) = \pi R^2$; and when $R-r < d < R+r$,
\begin{align}
    S(r,d) = & r^2 \arccos{\left(\frac{r^2 + R^2 - d^2}{2rR} \right)} + d^2 \arccos{\left(\frac{r^2 + d^2 - R^2}{2rd} \right)} \nonumber \\
    & \, \, \, - \frac{1}{2} \sqrt{(-r+R+D)(r+R-D)(r-R+d)(r+R+d)}.   
\end{align}

The probability density distribution for a baseline of length $d$ is the derivative of this CDF with respect to $d$:
\begin{align}
    \mathbb{P}(d) & = \frac{\partial \mathbb{P}(|\vec{A} - \vec{B}| \leq d)}{\partial d} = \frac{2}{\pi R^4} \int^R_0 r \frac{\partial S(r,d)}{\partial d} dr \nonumber \\
    & = \frac{4d}{\pi R^2} \left[\arccos{\left(\frac{d}{2R}\right)} - \frac{d}{2R} \sqrt{1 - \left(\frac{d}{2R} \right)^2}\right] \nonumber \\
    & = \frac{16 d}{\pi D_{\rm max}^2} \left[\arccos{\left(\frac{d}{D_{\rm max}}\right)} - \frac{d}{D_{\rm max}} \sqrt{1 - \left(\frac{d}{D_{\rm max}} \right)^2}\right].
    \label{eq:PDF_single_circular}
\end{align}
Therefore, the expected number of baselines of length $d$ per area in the $uv$‑plane is 
\begin{align}
    n_b(d) = \frac{N_{\rm tot} \mathbb{P}(d)}{2 \pi d} = \frac{N_{\rm ant}(N_{\rm ant}-1)}{4 \pi d} \mathbb{P}(d).
\end{align}

\subsection{Two Circular Arrays Separated by a Distance}\label{app:two_stations}
Suppose there are two circular arrays separated by a distance $L$, each with diameter $D$ and $N$ antennas. We derive the baseline density distribution for baselines formed between antennas from different arrays. 

Let $\vec{L}$ be the vector from the center of the first array $A$ to the center of the second array $B$. For simplicity, we place the first array at the origin $(0,0)$ and the second at $(L,0)$, so $\vec{L} = (L, 0)$. Consider a point in array $A$ at position $\vec{r}_A$ and a point in array $B$ at position $\vec{L} + \vec{r}_B$, where $\vec{r}_i$ are relative to their respective array centers. The baseline formed by these two antennas is $\vec{b} = \vec{L} + (\vec{r}_B - \vec{r}_A) = \vec{L} + \vec{r}$ with $\vec{r} \equiv \vec{r}_B - \vec{r}_A$. 

Since $\vec{b}$ is only a shift of $\vec{r}$, the probability distribution of $\vec{b}$ follows Eq.~\ref{eq:PDF_single_circular}, denoted here as $\mathbb{P}_{\rm intra}(d)$.
\begin{align}
    \mathbb{P}_{\rm cross}(\vec{b}) = \frac{1}{2 \pi |\vec{b} - \vec{L}|}\mathbb{P}_{\rm intra}(|\vec{b} - \vec{L}|).
\end{align}
Assuming isotropy of the baseline distribution function, we integrate over all angles and convert the vector $\vec{b}$ to its physical length $d$:
\begin{align}
    \mathbb{P}_{\rm cross}(d \equiv |\vec{b}|) = \int^{2 \pi}_0 |\vec{b}| \frac{1}{2 \pi |\vec{b} - \vec{L}|} \mathbb{P}_{\rm intra}(|\vec{b} - \vec{L}|) d\phi.
\end{align}
Note that only $|\vec{b} - \vec{L}|$ depends on $\phi$ and $|\vec{b} - \vec{L}| = \sqrt{d^2 + L^2 - 2dL \cos{\phi}}$. Thus, the probability distribution function becomes
\begin{align}
    \mathbb{P}_{\rm cross}(d) = d \int^{2\pi}_0 \frac{1}{2 \pi \sqrt{d^2 + L^2 - 2dL \cos{\phi}}}\mathbb{P}_{\rm intra}(\sqrt{d^2 + L^2 - 2dL \cos{\phi}}) d\phi.
    \label{eq:pdf_cross}
\end{align}
Therefore, the number density of baselines formed by antennas from different arrays is
\begin{align}
   n_{\rm cross}(d) = N^2 d \int^{2\pi}_0 \frac{\mathbb{P}_{\rm intra}(\sqrt{d^2 + L^2 - 2dL \cos{\phi}})}{2 \pi \sqrt{d^2 + L^2 - 2dL \cos{\phi}}} d\phi.
\end{align}

\bibliographystyle{JHEP} 
\bibliography{main}
\end{document}